\documentclass[article,nojss]{jss}

\usepackage{thumbpdf,lmodern}
\usepackage{orcidlink}

\usepackage{framed}

\usepackage{booktabs}
\usepackage{amsmath}
\usepackage{amssymb}
\usepackage{mathtools}
\usepackage{multirow}
\usepackage{dsfont}
\usepackage{algorithmic}
\usepackage{algorithm}

\newcommand{\fct}[1]{\code{#1()}}
\def\spacingset#1{\renewcommand{\baselinestretch}%
{#1}\small\normalsize} \spacingset{1}

\mathchardef\mhyphen="2D
\def \dsR {\text{$\mathds{R}$}}

\author{Nikolaus Umlauf~\orcidlink{0000-0003-2160-9803}\\Universit\"at Innsbruck
   \And Johannes Seiler~\orcidlink{0000-0001-5714-9234}\\Universit\"at Innsbruck
   \And Mattias Wetscher~\orcidlink{0000-0002-9982-3001}\\Universit\"at Innsbruck
   \AND Thorsten Simon~\orcidlink{0000-0002-3778-7738}\\Universit\"at Innsbruck
   \And Stefan Lang~\orcidlink{0000-0003-0739-3858}\\Universit\"at Innsbruck
   \And Nadja Klein~\orcidlink{0000-0002-5072-5347}\\Technische Universit\"at Dortmund
}
\Plainauthor{Umlauf~N., Seiler~J., Wetscher~M., Simon~T., Lang~S., Klein~N.}

\title{Scalable Estimation for Structured Additive Distributional Regression}
\Plaintitle{Scalable Estimation for Structured Additive Distributional Regression}
\Shorttitle{Scalable Estimation for Structured Additive Distributional Regression}

\Abstract{
  Recently, fitting probabilistic models have gained importance in many areas but estimation of
  such distributional models with very large data sets is a difficult task.
  In particular, the use of rather complex models can easily lead to memory-related efficiency
  problems that can make estimation infeasible even on high-performance computers.
  We therefore propose a novel backfitting algorithm,
  which is based on the ideas of stochastic gradient descent and can deal virtually with any
  amount of data on a conventional laptop. The algorithm performs automatic selection of
  variables and smoothing parameters, and its performance is in most cases superior or at
  least equivalent to other implementations for structured additive distributional regression,
  e.g., gradient boosting, while maintaining low computation time.
  Performance is evaluated using an extensive simulation study and an exceptionally challenging and 
  unique example of lightning count prediction over Austria. A very large dataset with over 9 
  million observations and 80 covariates is used, so that a prediction model cannot be estimated 
  with standard distributional regression methods but with our new approach.
}

\Plainkeywords{Generalized additive models for location, scale and shape; gradient descent; iteratively weighted least squares; stochastic optimization}

\Address{
  Nikolaus Umlauf, Johannes Seiler, Mattias Wetscher, Thorsten Simon, Stefan Lang\\
  Department of Statistics\\
  Faculty of Economics and Statistics\\
  Universit\"at Innsbruck\\
  Universit\"atsstr.~15\\
  6020 Innsbruck, Austria\\
  E-mail: \email{Nikolaus.Umlauf@uibk.ac.at},\\
  \phantom{E-mail: }\email{Johannes.Seiler@uibk.ac.at},\\
  \phantom{E-mail: }\email{Mattias.Wetscher@uibk.ac.at},\\
  \phantom{E-mail: }\email{Thorsten.Simon@uibk.ac.at},\\
  \phantom{E-mail: }\email{Stefan.Lang@uibk.ac.at}\\
  URL: \url{https://eeecon.uibk.ac.at/~umlauf/},\\
  \phantom{URL: }\url{https://www.uibk.ac.at/statistics/personal/lang/}\\

  Nadja Klein\\
  Chair of Uncertainty Quantification and Statistical Learning\\
  Research Center Trustworthy Data Science and Security (UA Ruhr)\\
  Department of Statistics (Technische Universit\"at Dortmund)\\
  Joseph-von-Fraunhofer-Str.~25\\
  44227 Dortmund, Germany\\
  E-mail: \email{nadja.klein@statistik.tu-dortmund.de}\\
  URL: \url{https://rc-trust.ai/klein/}
}

\begin{document}

\section{Introduction} \label{sec:introduction}

Fitting distributional regression models of high complexity
to large data is challenging with respect to storage and computational feasibility due to data
volume or very high-dimensional vectors of model parameters required to define sufficiently flexible
models. Moreover, in many applications, solving the problem also requires automatic selection of
variables since manual or stepwise searches in such model spaces are impossible to be conducted. In
recent years, techniques have already been developed to efficiently estimate generalized additive
models \citep[GAM;][]{sdr:Hastie+Tibshirani:1990,sdr:Fahrmeir+Kneib+Lang:2004} and generalized additive models for location
scale and shape \citep[GAMLSS;][]{sdr:Rigby+Stasinopoulos:2005,sdr:Klein+Kneib+Lang:2015}. For example,
\citet{sdr:Wood+Li+Shaddick:2017,sdr:Li+Wood:2019} show how to decompose the iterative
estimation algorithm for GAMs to be able to compute models for large data and gigadata with
coefficients up to $10^4$ and up to $10^8$ observations.
\citet{sdr:Lang+Umlauf+Wechselberger+Harttgen+Kneib:2014} present efficient algorithms for Bayesian
multilevel models for example by, discretization and indexing to significantly reduce the number of
floating point operations. These ideas are carried over to estimate fully Bayesian structured
additive distributional regression models \citep{sdr:Klein+Kneib+Lang+Sohn:2015}, the Bayesian version of GAMLSS,
such that e.g.~modelling the precipitation climatology across Austria with over $1.2$ million daily
observations is possible \citep{sdr:Umlauf+Klein+Zeileis:2018}. While in principle being easily
trainable in terms of data size with the approach of
\citet{sdr:Li+Wood:2019}, GAMs are not suited here given the censored nature of the response daily
precipitation with  a spike at zero. Nevertheless, for more complicated probabilistic models
or larger $n$, techniques such as \citet{sdr:Umlauf+Klein+Zeileis:2018}  also reach their limits.
On the one hand, such models can no longer be computed on conventional computers since there is
simply a lack of random-access memory (RAM); on the other hand, the computing time increases so
much that modeling with many variables is not possible in a foreseeable time.

To break down these barriers in structured additive distributional regression models, we propose a novel estimation
algorithm, which we call \emph{batchwise backfitting} and which combines the ideas of the classic
\textit{backfitting} optimization with \textit{stochastic gradient descent (SGD)}, an efficient
algorithm based on a stochastic approximation to gradient descent for finding local maxima of an
objective function $J(\boldsymbol{\theta})$ of a parameter vector $\boldsymbol{\theta} \in \dsR^p$
\citep{RobMon1951}. Compared to costly gradient descent methods, which involve updates of the form
$\boldsymbol{\theta} = \boldsymbol{\theta} - \boldsymbol{\eta}\nabla_\theta J(\boldsymbol{\theta})$
based on the whole data set, SGD replaces the gradient $\nabla_\theta J(\boldsymbol{\theta})$ by a
noisy (yet unbiased) estimate thereof, thus being much faster to compute. However, convergence to a
local optimum, which is theoretically guaranteed as long as the learning rate vector
$\boldsymbol{\eta}$ fulfils the Robbins-Monroe conditions
\citep{RobMon1951} can be extremely slow.

We show that our batchwise backfitting algorithm induces a learning rate that can be decomposed
into the product of a scalar step length $\nu$ and an adaptive learning rate vector
$\boldsymbol{\delta}$ based on second order information of the objective
function through an unbiased estimate of the Hessian, similar to the concept of natural gradients
motivated from information theory \citep{Ama1998, pmlr-v119-duan20a}.
The result is an algorithm that requires little manual tuning and ensures fast
convergence. Depending on the choice of $\nu$ we show that our algorithm closely mimics special
cases such as resampling or gradient boosting \citep{EfroTibs93, sdr:Mayr+Fenske+Hofner:2012}.
In addition, we demonstrate that our new algorithm
does not only significantly reduce computation time and requires extremely little memory, it also
has excellent properties in terms of variable selection; thus markedly contributing to a wider
applicability of structured additive distributional regression to \emph{big data} and
\emph{highly parameterized models}.

The remainder of the paper is structured as follows. In Section~\ref{sec:modelspec}, structured
additive distributional regression models are briefly reviewed. In Section~\ref{sec:estimation}, the new
\emph{batchwise backfitting} algorithm and its implementation for distributional regression models
is presented. In an extensive simulation study in Section~\ref{sec:simulation}, the performance of
the algorithm is investigated, whereas in Section~\ref{sec:lightapp} we further highlight the usefulness
of the algorithm developing a distributional model for lightning count forecasting using
a very large data set with $\approx 9.1$ million observations and 80 covariates. The final Section \ref{sec:summary} concludes.
Additional details on how to use our software implementation and
further simulation results are contained in the Appendix.

\pagebreak

\section{Structured Additive Distributional Regression Models} \label{sec:modelspec}

\subsection{Model Specification}
The idea in structured additive distributional regression
\citep[or GAMLSS;][]{sdr:Rigby+Stasinopoulos:2005, sdr:Klein+Kneib+Klasen+Lang:2015}
is to model all distributional parameters of an arbitrary parametric response distribution (rather
than just the mean) through covariates. Based on data $(Y_i,\mathbf{x}_i)$ of responses $Y_i$
(possibly non-continuous or multivariate, i.e.~$Y_i={\boldsymbol{Y}}_i\in\dsR^D$, $D>1$) and available covariate information
$\mathbf{x}_i$, for $i = 1, \ldots, n$ observations, we assume conditional independence of
individual response observations given covariates. Specifically
\begin{equation*} \label{eqn:dreg}
Y|\mathbf{x} \sim \mathbf{\mathcal{D}}_Y\left(\theta_1(\mathbf{x}))=h_{1}^{-1}(\eta_{1}(\mathbf{x})), \,
  \theta_2(\mathbf{x}))=h_{2}^{-1}(\eta_{2}(\mathbf{x})),\,\ldots\,\theta_K(\mathbf{x}))= h_{K}^{-1}(\eta_{K}(\mathbf{x}))\right)
\end{equation*}
where $\mathbf{\mathcal{D}_Y}$ denotes a parametric distribution with $K$ parameters
$\theta_k\equiv\theta_k(\mathbf{x})$, $k = 1, \ldots, K$, and parametric density $d_Y(\cdot;\theta_1,\ldots,\theta_K)$. Each parameter $\theta_k$ is  linked to an additive predictor
$\eta_k\equiv\eta_k(\mathbf{x})$ using known monotonic and twice differentiable functions
$h_{k}(\cdot)$ (with inverses $h_k^{-1}$ also known as link- and response functions) to ensure potential parameter space restrictions on $\theta_k$. The additive predictor for the $k$-th parameter is modeled as
\begin{equation} \label{eqn:addpred}
{\eta}_k = \eta_k(\mathbf{x}; \boldsymbol{\beta}_k) =
  f_{1k}(\mathbf{x}; \boldsymbol{\beta}_{1k}) + \ldots + f_{J_kk}(\mathbf{x}; \boldsymbol{\beta}_{J_kk}),
\end{equation}
based on $j = 1, \ldots, J_k$ unspecified (possibly non-linear) functions $f_{jk}(\cdot)$, applied
to a subset of $\mathbf{x}$. For a data set of $i=1,\ldots,n$ observations, let $\mathbf{X}$ be the
covariate matrix with rows $\mathbf{x}_i$, and
$\boldsymbol{\eta}_k=(\eta_{1,k},\ldots,\eta_{n,k})^\top$ be
the corresponding $n$ dimensional vector of predictors each entry containing the sum of evaluations of
$f_{jk}( \cdot )$ at $\mathbf{x}_i$. The parameters
$\boldsymbol{\beta}_k = (\boldsymbol{\beta}_{1k}, \ldots, \boldsymbol{\beta}_{J_kk})^\top$ are
the regression coefficients and we denote furthermore
$\mathbf{X}_k = (\mathbf{X}_{1k}, \ldots, \mathbf{X}_{J_kk})$ the predictor specific design
matrices, whose structure only depend on the type of covariate(s) and assumptions about
$f_{jk}( \cdot )$. For the models discussed here, matrices $\mathbf{X}_{jk}$ are typically based on
a basis function approach, e.g., using B-spline basis functions \citep{sdr:Eilers+Marx:1996} or
thin-plate splines \citep{Wood:2003} for modeling smooth effects. Therefore, each function
$f_{jk}(\cdot)$ may be represented by the linear combination
$f_{jk}(\mathbf{X}_{jk}, \boldsymbol{\beta}_{jk}) = \mathbf{X}_{jk}\boldsymbol{\beta}_{jk}$
which leads to so-called GAM-type or structured additive predictors
$\boldsymbol{\eta}_k$~\citep[STAR,][]{sdr:Fahrmeir+Kneib+Lang:2004}.

\subsection{Penalized Likelihood Estimation}
Likelihood-based estimation in this flexible model class is typically based on the penalized log-likelihood function
\begin{equation} \label{eqn:logpost}
  \ell_{\text{pen}}(\boldsymbol{\beta}, \boldsymbol{\tau} ; \mathbf{y}, \mathbf{X}) =
  \ell(\boldsymbol{\beta} ; \mathbf{y}, \mathbf{X}) +
  \sum_{k = 1}^K\sum_{j = 1}^{J_k} P_{jk}(\boldsymbol{\beta}_{jk},\boldsymbol{\tau}_{jk}),
\end{equation}
where $\ell(\boldsymbol{\beta} ; \mathbf{y}, \mathbf{X})$ is log-likelihood function
\begin{equation*} \label{eqn:loglik}
\ell(\boldsymbol{\beta} ; \mathbf{y}, \mathbf{X}) =
  \sum_{i = 1}^n \log \, d_Y(y_i ; \theta_{i,1} = h_1^{-1}(\eta_{1}(\mathbf{x}_i; \boldsymbol{\beta}_1)), \ldots,
  \theta_{i,K} = h_K^{-1}(\eta_{K}(\mathbf{x}_i; \boldsymbol{\beta}_K))),
\end{equation*}
$\boldsymbol{\theta}_k = (\theta_{1,k}, \ldots, \theta_ {n,k})^\top$ are the parameter vectors and $\boldsymbol{\beta} = (\boldsymbol{\beta}_1^\top, \ldots, \boldsymbol{\beta}_K^\top)^\top$ the
stacked vector of regression coefficients
 to be
estimated. The overall design matrix is
$\mathbf{X} = (\mathbf{X}_1, \ldots, \mathbf{X}_K)$, where each $\mathbf{X}_k$ consists of rows $\mathbf{x}_{i,k}$. To avoid the problem of overfitting,  each
function $f_{jk}(\cdot)$ is regularized through the penalty terms $P_{jk}(\boldsymbol{\beta}_{jk},\boldsymbol{\tau}_{jk})$, where $\boldsymbol{\tau}_{jk}$ controls the amount of smoothness and $P_{jk}( \cdot )$ is specific to $f_{jk}(\cdot)$. In
general, the penalty terms are assumed to be of the following quadratic form
\begin{equation} \label{eqn:classicpen}
P_{jk}(\boldsymbol{\beta}_{jk},\boldsymbol{\tau}_{jk}) =
  \boldsymbol{\beta}_{jk}^\top\mathbf{K}_{jk}(\boldsymbol{\tau}_{jk})\boldsymbol{\beta}_{jk}.
\end{equation}
For instance, when using P-splines, $P_{jk}( \cdot )$ is computed by a penalty matrix
$\boldsymbol{\tau}_{jk}\mathbf{K}_{jk}$ formed by the cross-product of difference matrices. This then penalizes too
abrupt jumps of neighboring coefficients to achieve a smooth functional form
\citep[a similar penalty structure results from, e.g., thin-plate splines or
tensor splines;][]{sdr:Fahrmeir+Kneib+Lang+Marx:2013,sdr:Wood:2017}.
\citet{sdr:Groll+Hambuckers+Kneib+Umlauf:2019} extend the classical smoothing penalty for GAMLSS
to (fused) LASSO-type penalties $P_{jk}(\boldsymbol{\beta}_{jk},\boldsymbol{\tau}_{jk}) =
  \boldsymbol{\beta}_{jk}^\top \mathbf{K}_{jk}(\boldsymbol{\tau}_{jk})\boldsymbol{\beta}_{jk}$,
where the penalty $\mathbf{K}_{jk}( \cdot )$ is also a function of the regression coefficients accounting
for (approximate) L$_1$-regularization \citep{sdr:Tibshirani:2005,sdr:OelTutz:2017}. In the
following, we will describe the algorithms with the ``classic'' penalization \eqref{eqn:classicpen}
for the sake of simplicity, but more complex penalties can be implemented just as straightforwardly.

\subsection{Backfitting}

To maximize \eqref{eqn:logpost}, \citet{sdr:Rigby+Stasinopoulos:2005} proposed a modified backfitting algorithm  based on
iteratively reweighted (penalized) least squares \citep[IRPLS;][]{Mar1996}, which similar to the
backfitting algorithm of \citet{sdr:Umlauf+Klein+Zeileis:2018} employs updates based on
iteratively weighted least squares \citep[IWLS;][]{sdr:Gamerman:1997}. The updating equation for
the $jk$-th model term of \eqref{eqn:addpred} is given by
\begin{equation} \label{eqn:bfit}
  \boldsymbol{\beta}_{jk}^{[t+1]} =
    (\mathbf{X}_{jk}^\top\mathbf{W}_{kk}\mathbf{X}_{jk} +
      \mathbf{K}_{jk}(\boldsymbol{\tau}_{jk}))^{-1}\mathbf{X}_{jk}^\top\mathbf{W}_{kk}(
      \mathbf{z}_k - \boldsymbol{\eta}_{k, -j}^{[t+1]}),
\end{equation}
with vector of working observations
$\mathbf{z}_k = \boldsymbol{\eta}_{k}^{[t]} + \mathbf{W}_{kk}^{-1}\mathbf{u}_k$, score vectors
$\mathbf{u}_k = \partial \ell(\boldsymbol{\beta}; \mathbf{y},
  \mathbf{X}) / \partial \boldsymbol{\eta}_k$ and 
working weights
$\mathbf{W}_{kk} = -\mathrm{diag}(\partial^2 \ell(\boldsymbol{\beta}; \mathbf{y}, \mathbf{X})
\partial \boldsymbol{\eta}_k \partial \boldsymbol{\eta}_k^\top)$. Here, $\boldsymbol{\eta}_{k, -j}$ represents the
predictor without the $j$-th model term. The backfitting iterations at \eqref{eqn:bfit}
are computed until a certain termination criterion is met, e.g., when the relative change of
the coefficients becomes very small. The optimal smoothing parameters can be estimated using e.g.~stepwise selection \citep{sdr:Belitz+Lang:2008}, where in each updating step at \eqref{eqn:bfit}
each $\boldsymbol{\tau}_{jk} = (\tau_{1jk}, \ldots, \tau_{L_{jk}jk})^\top$ is optimized one after
the other using adaptive search intervals, e.g., using the Akaike (AIC) or Bayesian information
criterion (BIC), noting that in many cases, $\boldsymbol{\tau}_{jk}$ is just a scalar. For a detailed
description of the algorithm see \citet{sdr:Umlauf+Klein+Zeileis:2018}. Moreover, for numerical
reasons it is oftentimes better to replace the Hessian by the expected Fisher information with
weights
$\mathbf{W}_{kk} = -\mathrm{diag}(E(\partial^2 \ell(\boldsymbol{\beta}; \mathbf{y}, \mathbf{X}) /
\partial \boldsymbol{\eta}_k \partial \boldsymbol{\eta}_k^\top))$
\citep[][]{sdr:Klein+Kneib+Lang:2015}. To  reduce computation times, the design matrix
$\mathbf{X}_{jk}$ can be modified by using only the unique values of the covariate data, which in
many cases have much less observations than the number of observations in the whole data set.
This leads to an updating step with reduced working observations and weights, which can be
calculated quickly via a simple sum with indices of the unique values
\citep[][]{sdr:Lang+Umlauf+Wechselberger+Harttgen+Kneib:2014}.
Although this method can save quite a bit of computing time, memory issues can still occur very
quickly in the GAMLSS model class.

\pagebreak

\section{Scalable Estimation} \label{sec:estimation}

As a solution to large-scale data, we present our batchwise
backfitting algorithm as part of this section first. Then we discuss some interesting properties of our algorithm depending on the step length choice but also further computational details. 

\subsection{Batchwise Backfitting}\label{sec:bbfit}

Instead of using all observations of the data, we replace score vector and Hessian in
\eqref{eqn:bfit} through unbiased estimates thereof, which are readily available based on a random
batch of the data. That is, we use a randomly chosen subset denoted by the subindex $[\mathbf{i}]\subseteq\lbrace 1,\ldots,n\rbrace$
to arrive at a stochastic updating step of the form
\begin{eqnarray} \label{eqn:bbfit}
\boldsymbol{\beta}_{jk}^{[t+1]} &=& (1 - \nu) \cdot \boldsymbol{\beta}_{jk}^{[t]} + \\
&& \hspace*{-1.7cm}\nu \cdot (\mathbf{X}_{[\mathbf{i}], jk}^\top\mathbf{W}_{[\mathbf{i}], kk}\mathbf{X}_{[\mathbf{i}], jk} +  \mathbf{K}_{jk}(\boldsymbol{\tau}_{jk}))^{-1}\mathbf{X}_{[\mathbf{i}], jk}^\top\mathbf{W}_{[\mathbf{i}], kk}(
\mathbf{z}_{[\mathbf{i}], k} - \boldsymbol{\eta}_{[\mathbf{i}], k, -j}^{[t+1]}) \nonumber \\
&=& (1 - \nu) \cdot \boldsymbol{\beta}_{jk}^{[t]} + \nu \cdot \boldsymbol{\beta}_{[\mathbf{i}], jk}
\nonumber
\end{eqnarray}
and introduce a step length control parameter $\nu$ (or \textit{learning rate}) specifying the
amount of which $\boldsymbol{\beta}_{jk}^{[t]}$ is updated to $\boldsymbol{\beta}_{jk}^{[t+1]}$ in the direction of the new estimate
$\boldsymbol{\beta}_{[\mathbf{i}], jk}$ on batch $[\mathbf{i}]$. In each iteration, \eqref{eqn:bbfit} is evaluated on exactly
one batch $[\mathbf{i}]$, such that computational burden can be reduced considerably . As mentioned in the introduction,
this mimics a second order SGD algorithm \citep{Bot2012} since
\begin{equation} \label{eqn:update}
\boldsymbol{\beta}^{[t+1]}_{jk} = \boldsymbol{\beta}^{[t]}_{jk} +
  \nu \cdot (\boldsymbol{\beta}_{[\mathbf{i}], jk} - \boldsymbol{\beta}^{[t]}_{jk}) =
  \boldsymbol{\beta}^{[t]}_{jk} + \nu \cdot \boldsymbol{\delta}^{[t]}_{jk},
\end{equation}
where the difference $\boldsymbol{\delta}^{[t]}_{jk}$ between parameter updates from iteration $t$
and batch $[\mathbf{i}]$ is a decomposition of first and second order derivative information with
\begin{eqnarray*}
\boldsymbol{\delta}^{[t]}_{jk} &=& \boldsymbol{\beta}_{[\mathbf{i}], jk} - \boldsymbol{\beta}^{[t]}_{jk} \\
  &=&  \left[\boldsymbol{\beta}^{[t]}_{jk} -
    \mathbf{H}_{[\mathbf{i}],kk}\left(\boldsymbol{\beta}^{[t]}_{jk}\right)^{-1}\mathbf{s}_{[\mathbf{i}]}\left(\boldsymbol{\beta}^{[t]}_{jk}\right)\right] - \boldsymbol{\beta}^{[t]}_{jk} \\
  &=& -\mathbf{H}_{[\mathbf{i}],kk}\left(\boldsymbol{\beta}^{[t]}_{jk}\right)^{-1}\mathbf{s}_{[\mathbf{i}]}\left(\boldsymbol{\beta}^{[t]}_{jk}\right),
\end{eqnarray*}
where $\mathbf{s}_{[\mathbf{i}]}( \cdot )$ and $\mathbf{H}_{[\mathbf{i}],kk}( \cdot )$ are unbiased
estimates of the score and Hessian (see also \citealp{sdr:Umlauf+Klein+Zeileis:2018})
evaluated on batch $[\mathbf{i}]$
\begin{eqnarray*} \label{eqn:score}
\mathbf{s}_{[\mathbf{i}]}(\boldsymbol{\beta}_{jk}) & = &
  \frac{\partial \ell_{\text{pen}}(\boldsymbol{\beta}, \boldsymbol{\tau} ; \mathbf{y}_{[\mathbf{i}]}, \mathbf{X}_{[\mathbf{i}]})}{
    \partial \boldsymbol{\beta}_{jk}}
= \frac{\partial \ell(\boldsymbol{\beta}; \mathbf{y}_{[\mathbf{i}]}, \mathbf{X}_{[\mathbf{i}]})}{\partial \boldsymbol{\beta}_{jk}} +
    \sum_{k = 1}^K\sum_{j = 1}^{J_k} \left[ \frac{\partial P(\boldsymbol{\beta}_{jk},\boldsymbol{\tau}_{jk})}{\partial \boldsymbol{\beta}_{jk}} \right],\\
\label{eqn:hessian}
\mathbf{H}_{[\mathbf{i}], kk}(\boldsymbol{\beta}_{jk}) & = &
\frac{\partial \mathbf{s}_{[\mathbf{i}]}(\boldsymbol{\beta}_{jk})}{\partial \boldsymbol{\beta}_{jk}^\top} =
\frac{\partial^2 \ell_{\text{pen}}(\boldsymbol{\beta}, \boldsymbol{\tau} ; \mathbf{y}_{[\mathbf{i}]}, \mathbf{X}_{[\mathbf{i}]})}{
  \partial \boldsymbol{\beta}_{jk} \partial \boldsymbol{\beta}_{jk}^\top} \\
  &=&\frac{\partial^2 \ell(\boldsymbol{\beta}, \boldsymbol{\tau} ; \mathbf{y}_{[\mathbf{i}]}, \mathbf{X}_{[\mathbf{i}]})}{
  \partial \boldsymbol{\beta}_{jk} \partial \boldsymbol{\beta}_{jk}^\top} +
    \sum_{k = 1}^K\sum_{j = 1}^{J_k} \left[ \frac{\partial P(\boldsymbol{\beta}_{jk},\boldsymbol{\tau}_{jk})}{\partial \boldsymbol{\beta}_{jk}\partial\boldsymbol{\beta}_{jk}^\top} \right].
\end{eqnarray*}
Using second order information can speed up convergence considerably and our updating rule
resembles that of natural gradients \citep{Ama1998}.
In each iteration of the batchwise backfitting algorithm the update step length is adaptive, because
of the curvature information provided in $\boldsymbol{\delta}^{[t]}_{jk}$.

The working weights $\mathbf{W}_{[\mathbf{i}], kk}$, the working responses
$\mathbf{z}_{[\mathbf{i}], k}$ and the predictors
$\boldsymbol{\eta}_{[\mathbf{i}], k}$ are computed based on the current states
$\boldsymbol{\beta}_k^{[t]}$. For each batch $[\mathbf{i}]$, the algorithm subsequently cycles over all
parameters of the response distribution, the outer loop, and all model terms, the inner loop,
in the typical backfitting manner, i.e.,
the predictors $\boldsymbol{\eta}_{[\mathbf{i}], k}$ and model terms $f_{jk}( \cdot )$ are updated
instantly within the inner loop. By iteration through the batches the batchwise
backfitting algorithm updates in a memory efficient manner from batch to batch either until all
observations are included once, or the algorithm runs through the data a prespecified number of
\textit{epochs}. This design principle makes the batchwise backfitting optimizer computationally
simple and thus \textit{scaleable}.

\subsection{Choosing the Batch Size}
The size of the batches is application specific. In general, a good strategy is to first estimate
intercept only models, with $\eta_k = \beta_{0k}$,  using batchwise backfitting and 
small batches, e.g., about 1000 observations, and then inspect the coefficient paths. If these are
stationary after a certain runtime, the batch size is sufficient and if not it should be increased
successively. For examples of coefficient paths that are stationary after a certain ``burn-in''
phase, see Figure~\ref{fig:parpaths}. This approach has proven successful, e.g., in the
application Section~\ref{sec:lightapp}.

\subsection{Choosing the Step Length}

Our default batchwise backfitting works with a fixed step length $\nu = 0.1$, which is a good
compromise between fast updates and numerical stability and has also been shown to be very robust
in simulations. In addition, we consider the following two variants of the basic algorithm.

\paragraph{Resampling Variant}
If $\nu = 1$, the algorithm can be interpreted as a \textit{resampling} method and each update
$\boldsymbol{\beta}_{jk}^{[t+1]}$ resembles a ``sample'' of the  ``distribution'' of $\boldsymbol{\beta}_{jk}$, and convergence is achieved in distribution, i.e., once the estimates are fluctuating around a
certain level. The final estimate $\hat{\boldsymbol{\beta}}$ is then computed by taking the means
or medians of the resulting coefficient paths after convergence.

\paragraph{Boosting Variant}
In addition, \eqref{eqn:update} can also be utilized to enforce complete variable selection in a
boosting type algorithm when only the model term with the best improvement in the
\textit{out-of-sample} log-likelihood is updated. An important innovation of this variant over
classical gradient boosting for GAMLSS is that the smoothing parameters $\boldsymbol{\tau}_{jk}$
are also updated in each iteration (see Section~\ref{sec:hyper}), i.e.,
the last iteration already leads to the final model.
In contrast, in classical boosting for GAMLSS the optimal stopping iteration is crucial and has to
be determined separately
\citep[commonly based on costly cross validation (CV); see, e.g.,][]{sdr:Mayr+Fenske+Hofner:2012, sdr:Thomas+Mayr+Bischl:2018}.
The costly CV makes boosting GAMLSS infeasible for big data. A further considerable advantage of our
algorithm is that it makes the selection of the best model term relatively fair, unlike boosting
variants with fixed prechosen degrees of freedom for $f_{jk}( \cdot )$. For example, with more
complicated distributions, it can easily happen that certain parameters are never selected because
of too large differences in the gradients. \citet{sdr:2102.09248} try to circumvent this problem
by adaptive step length selection for $\nu$ in the linear normal location-scale model,
however, for the general class of GAMLSS this procedure seems to be difficult or even impossible
to implement.

\begin{figure}[t!]
\centering
\includegraphics[width=1\textwidth]{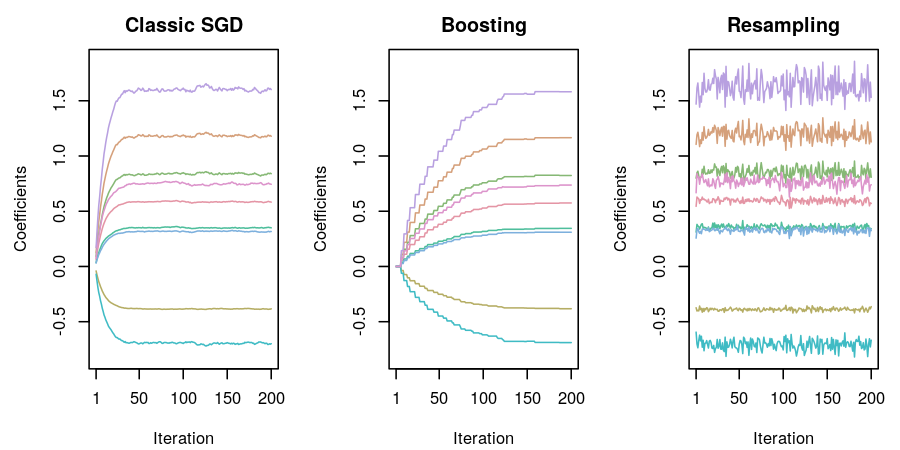}
\caption{\label{fig:parpaths} Examples of coefficient paths for $\boldsymbol{\beta}_{jk}$
  of a spline model term $f_{jk}( \cdot )$ using the three different variants of the
  batchwise backfitting algorithm.}
\end{figure}

\paragraph{Graphical Illustration}
The three different variants of the algorithm are illustrated in Figure~\ref{fig:parpaths}.
Here, the coefficient paths are shown for a model term estimated with a thin-plate spline. The
left plot shows coefficient paths of the batchwise backfitting with $\nu = 0.1$, it takes
approximately $50$ iterations for the coefficients to reach a steady state. The middle plot
illustrates coefficients paths for the boosting version with $\nu = 0.1$ of the algorithm and
possible updating only if the relative improvement of the log-likelihood on the next batch
$[\tilde{\mathbf{i}}]$ is larger than a prespecified constant $c$. In the first few iterations,
the model term is not selected, all coefficients are zero. Around iteration $10$, coefficients
start to deviate from zero and converge to a steady state shortly after iteration $150$. After
that, the coefficients are no longer updated, as indicated by the strict horizontal movements. The
right plot shows coefficient paths if the step length is set to $\nu = 1$ and updates are always
allowed in combination with slice sampling of the smoothing parameters $\boldsymbol{\tau}_{jk}$
under the AIC using the next batch $[\tilde{\mathbf{i}}]$. Similar to the basic batchwise
backfitting algorithm, the coefficients require about $50$ iterations to reach a steady state.
randomly from a proposal density and an acceptance step is not required. 

\subsection{Estimation of Hyperparameters} \label{sec:hyper}

As described in Section~\ref{eqn:loglik}, the smoothness of $f_{jk}( \cdot )$ is controlled by
parameters $\boldsymbol{\tau}_{jk}$. In the proposed implementation these parameters are either
estimated according to an information criterion like the AIC or BIC, which is computed on an
\textit{out-of-sample} batch $[\tilde{\mathbf{i}}]$, or by slice sampling under the information
criterion \citep{sdr:Neal:2003}. Using the \textit{out-of-sample} batch for selection is a novelty,
aiming to improve the predictive performance of the model. Moreover, in addition to commonly used
penalties in $P_{jk}( \cdot )$, complete model term selection can also be incorporated by an
additional LASSO-type penalty for coefficients $\boldsymbol{\beta}_{jk}$
\citep{sdr:Groll+Hambuckers+Kneib+Umlauf:2019}.

\subsection{Computational Details and Implementation}

The complete algorithm is described in pseudo code in Algorithm~\ref{fig:algodesign} and is
implemented in the \proglang{R} package \pkg{bamlss} \citep{sdr:Umlauf+Klein+Zeileis:2016}
within the optimizer function \code{opt\_bbfit()}.
It supports all commonly used model terms for GAMs, as implemented in the
\pkg{mgcv} package \citep{sdr:Wood:2019}.
In addition, to overcome memory issues with very large data, the \pkg{bamlss} package now supports
the binary flat file format for data frames, which is implemented in the \pkg{ff} package
\citep{sdr:ff}. By processing data and design matrices with \pkg{ff}, the usual memory limitations
of the \proglang{R} ecosystem are circumvented.
This is achieved by loading the data sequentially, using chunks that fit in memory, so that the
complete data is never in the RAM. This means that the batchwise backfitting optimizer
\code{opt\_bbfit()} can work directly with \pkg{ff} objects, i.e., the batches are loaded directly
by the \pkg{ff} infrastructure, which usually means only very little additional processing time.
This makes it possible to use almost arbitrarily large data sets for the estimation of structured additive distributional
regression models. In Appendix~\ref{appendix:estimation}, we give detailed examples on how to
fit models with the new optimizer function and its handling within the \pkg{bamlss} framework
using simulated data with $10^7$ observations.

\begin{algorithm}[p!]
\spacingset{1}
\caption{\label{fig:algodesign} Batchwise backfitting.}
\small
{\fontsize{10}{5}
\begin{algorithmic}
\renewcommand{\algorithmicrequire}{\textbf{Input:}}
\medskip
\REQUIRE{$\mathbf{y}$, $\mathbf{X}$, $\boldsymbol{\alpha}$.}
\renewcommand{\algorithmicrequire}{\textbf{Initialize:}}
\renewcommand{\algorithmicensure}{\textbf{Set:}}
\ENSURE{Step length $\nu \in [0, 1]$,
   batch index $\mathbf{B} = (\mathbf{b}_1, \ldots, \mathbf{b}_T)^\top$,
   goodness-of-fit criterion $C$, \newline
   scaling constant $c \in \mathbb{R}$, e.g., $c = 1$.}
\REQUIRE{$\boldsymbol{\beta}$, $\boldsymbol{\tau}$, e.g.,
  $\boldsymbol{\beta} = \mathbf{0}, \boldsymbol{\tau} = 0.001 \cdot \mathbf{1}$.}
\medskip
\FOR{$t$ in $1$ to number of batches $T - 1$.}
  \STATE Set current batch index $\mathbf{i} = \mathbf{b}_t$.
  \STATE Set next batch index, e.g., with $\tilde{\mathbf{i}} = \mathbf{b}_{t + 1}$.
  \medskip
  \FOR{$k = 1$ to $K$}
    \STATE Initialize $\boldsymbol{\eta}_{[\mathbf{i}], k} = \mathbf{0}$.
    \FOR{$j = 1$ to $J_k$}
      \STATE Compute state $\boldsymbol{\eta}_{[\mathbf{i}], k} = \boldsymbol{\eta}_{[\mathbf{i}], k} +
        \mathbf{X}_{[\mathbf{i}], jk}\boldsymbol{\beta}_{jk}^{[t]}$ on current batch.
    \ENDFOR
  \ENDFOR
  \STATE Likewise compute $\boldsymbol{\eta}_{[\tilde{\mathbf{i}}], k}$ on next batch.
  \medskip
  \FOR{$k = 1$ to $K$}
    \FOR{$j = 1$ to $J_k$}
      \STATE Compute old log-likelihood on next batch $\ell(\boldsymbol{\beta}^{[t]}; \mathbf{y}_{[\tilde{\mathbf{i}}]}, \mathbf{X}_{[\tilde{\mathbf{i}}]})$.
      \STATE Set the working response $\mathbf{z}_{[\mathbf{i}], k} = \boldsymbol{\eta}_{[\mathbf{i}], k} +
        \mathbf{W}_{[\mathbf{i}], kk}^{-1}\mathbf{u}_{[\mathbf{i}], k}$.
      \STATE IWLS step \newline
        $\phantom{ii}\boldsymbol{\beta}_{[\mathbf{i}], jk} =
        (\mathbf{X}_{[\mathbf{i}], jk}^\top\mathbf{W}_{[\mathbf{i}], kk}\mathbf{X}_{[\mathbf{i}], jk} +
        \mathbf{K}_{jk}))^{-1}\mathbf{X}_{[\mathbf{i}], jk}^\top\mathbf{W}_{[\mathbf{i}], kk}(\mathbf{z}_{[\mathbf{i}], k} - \boldsymbol{\eta}_{[\mathbf{i}], -j,k})$.
      \STATE Therefore find new $\boldsymbol{\tau}_{jk}^{[t + 1]}$ on next batch.
      \FOR{$l = 1$ to $L_{jk}$}
        \STATE Set search interval for $\tau_{ljk}^{[t + 1]}$, e.g., $\mathcal{I}_{ljk} = [\tau_{ljk}^{[t]} \cdot 10^{-1}, \tau_{ljk}^{[t]} \cdot 10]$.
        \STATE Find $\tau_{ljk}^{[t + 1]} \leftarrow
    \underset{\tau_{ljk}^\star \in \mathcal{I}_{ljk}}{\text{arg min }} C(\boldsymbol{\beta}_{[\mathbf{i}], jk}, \tau_{ljk}^\star; \mathbf{y}_{[\tilde{\mathbf{i}}]}, \mathbf{X}_{[\tilde{\mathbf{i}}]})$,
        or slice sample under $C( \cdot )$.
      \ENDFOR
      \medskip
      \STATE Now set $\mathring{\boldsymbol{\beta}}_{jk} = \boldsymbol{\beta}^{[t]}_{jk} +
        \nu \cdot (\boldsymbol{\beta}_{[\mathbf{i}], jk} - \boldsymbol{\beta}^{[t]}_{jk})$.

      \IF{Updated log-likelihood $\ell(\mathring{\boldsymbol{\beta}}; \mathbf{y}_{[\tilde{\mathbf{i}}]}, \mathbf{X}_{[\tilde{\mathbf{i}}]}) > c \cdot \ell(\boldsymbol{\beta}^{[t]}; \mathbf{y}_{[\tilde{\mathbf{i}}]}, \mathbf{X}_{[\tilde{\mathbf{i}}]})$.}
        \STATE Update $\boldsymbol{\beta}_{jk}^{[t + 1]} = \mathring{\boldsymbol{\beta}}_{jk}$.
        \STATE Update $\boldsymbol{\eta}_{[\mathbf{i}], k} = \boldsymbol{\eta}_{[\mathbf{i}], k} +
        \mathbf{X}_{[\mathbf{i}], jk}\boldsymbol{\beta}_{jk}^{[t + 1]}$ and likewise $\boldsymbol{\eta}_{[\tilde{\mathbf{i}}], k}$.
      \ELSE
        \STATE Set $\boldsymbol{\beta}_{jk}^{[t + 1]} = \boldsymbol{\beta}_{jk}^{[t]}$.
      \ENDIF
    \ENDFOR
  \ENDFOR
  \medskip
  \STATE Alternatively, only update coefficients $\boldsymbol{\beta}_{jk}$ which lead to the
    greatest contribution in the next batch log-likelihood.
  \medskip
\ENDFOR
\medskip
\renewcommand{\algorithmicensure}{\textbf{Output:}}
\ENSURE{Estimates $\hat{\boldsymbol{\beta}} = \boldsymbol{\beta}^{[T - 1]}$;
  or ``samples'' $\boldsymbol{\beta}^{[t]}$, $t = 1, \ldots, T - 1$;
  or boosting like coefficient paths $\boldsymbol{\beta}^{[t]}$, $t = 1, \ldots, T - 1$
  if only the best working model term is updated in each batch.}
\medskip
\end{algorithmic}
}
\end{algorithm}

\section{Simulation Study} \label{sec:simulation}

To investigate the performance of the proposed batchwise backfitting algorithm in terms of variable
selection, mean squared error (MSE), prediction and runtimes, we conduct a benchmark study against
classical Markov chain Monte Carlo (MCMC) and gradient boosting algorithms for GAMLSS for which we
give details next before describing the simulation design and results.

\subsection{Estimation Approaches}

In the following, we refer to our proposed approach of batchwise backfitting throughout as 
\code{opt\_bbfit} (as the model fitting function is called in the \pkg{bamlss} package).
Our batchwise backfitting  combines the boosting and the resampling variant as described
in Section~\ref{sec:bbfit}. The boosting step is run for $400$ iterations including all possible
covariates. This first step is used to preselect the covariates, and only covariates that are
updated at least once are included in the subsequent resampling variant of the algorithm, which is
run for $1500$ iterations. The batch indices are drawn randomly, for the very small datasets
of $500$ observations we use a batchsize of $400$, for larger datasets up to $10000$ observations 
the batchsize is 63\% of the data, for settings with $\ge 10000$ observations, the batchsize is 
fixed constant at $10000$.

We investigate the performance of our batchwise backfitting \code{opt\_bbfit} approach
compared to the following very popular methods in distributional regression.
\begin{enumerate}
\item \textit{MCMC} (\code{sam\_mcmc}).
  The default MCMC implementation of the \pkg{bamlss} package \citep{sdr:Umlauf+Klein+Zeileis:2016}
  in \proglang{R} based on IWLS proposals is used.
  Note that the \code{sam\_mcmc} method does not perform variable selection and therefore
  serves as an unconstrained benchmark. 

\item \textit{Non-Cyclical Gradient Boosting} (\code{gamboostLSS}).
  Gradient boosting for GAMLSS combines an ensemble of weak base learners. Instead of updating
  every distributional parameter with a base learner in each iteration (cyclical),
  in the non-cyclical gradient boosting version \citep{sdr:Thomas+Mayr+Bischl:2018} the
  algorithm updates only the base learner (model term) which leads to the highest loss
  reduction over all distributional parameters in every iteration.
  The intercepts are always updated. The optimal stopping iteration (\textit{mstop}) is selected
  by five-fold CV. The non-cyclical gradient boosting algorithm is implemented
  in the \proglang{R} package \pkg{gamboostLSS} \citep{sdr:Hofner2021}.

\item \textit{Optimized Non-Cyclical Gradient Boosting} (\code{opt\_boost}).
  The optimized version of the non-cyclic gradient boosting algorithm is implemented in
  the \proglang{R} package \pkg{bamlss} and utilizes methods for large data sets,
  originally designed to achieve speed improvements in MCMC algorithms 
  \citep{sdr:Lang+Umlauf+Wechselberger+Harttgen+Kneib:2014}. Unlike the classical
  non-cyclic gradient boosting algorithm, the model intercepts count as single model terms
  and are not automatically updated. Five-fold CV is applied to find the optimal stopping
  iteration.
\end{enumerate}

\subsection{Simulation Design}

\paragraph{Response Distributions}
We simulate data from the normal distribution (\code{NO}), the gamma distribution (\code{GA}), and
the zero-adjusted Poisson distribution (\code{ZAP}). All three distributions are implemented in the
\proglang{R} package \pkg{gamlss.dist} \citep{sdr:gamlss.dist}. The package uses a specific naming
convention for the parameters of the distributions, supporting up to four-parameter distributions.
The parameters are $\mu$, $\sigma$, $\nu$ and $\tau$. In the simulation study, we let parameters
$\mu$ and $\sigma$ depend on covariates. Since all distributions studied in this setting have
two parameters, no specifications for $\nu$ and $\tau$ are needed.

\begin{figure}[t!]
\centering
\includegraphics[width=0.9\textwidth]{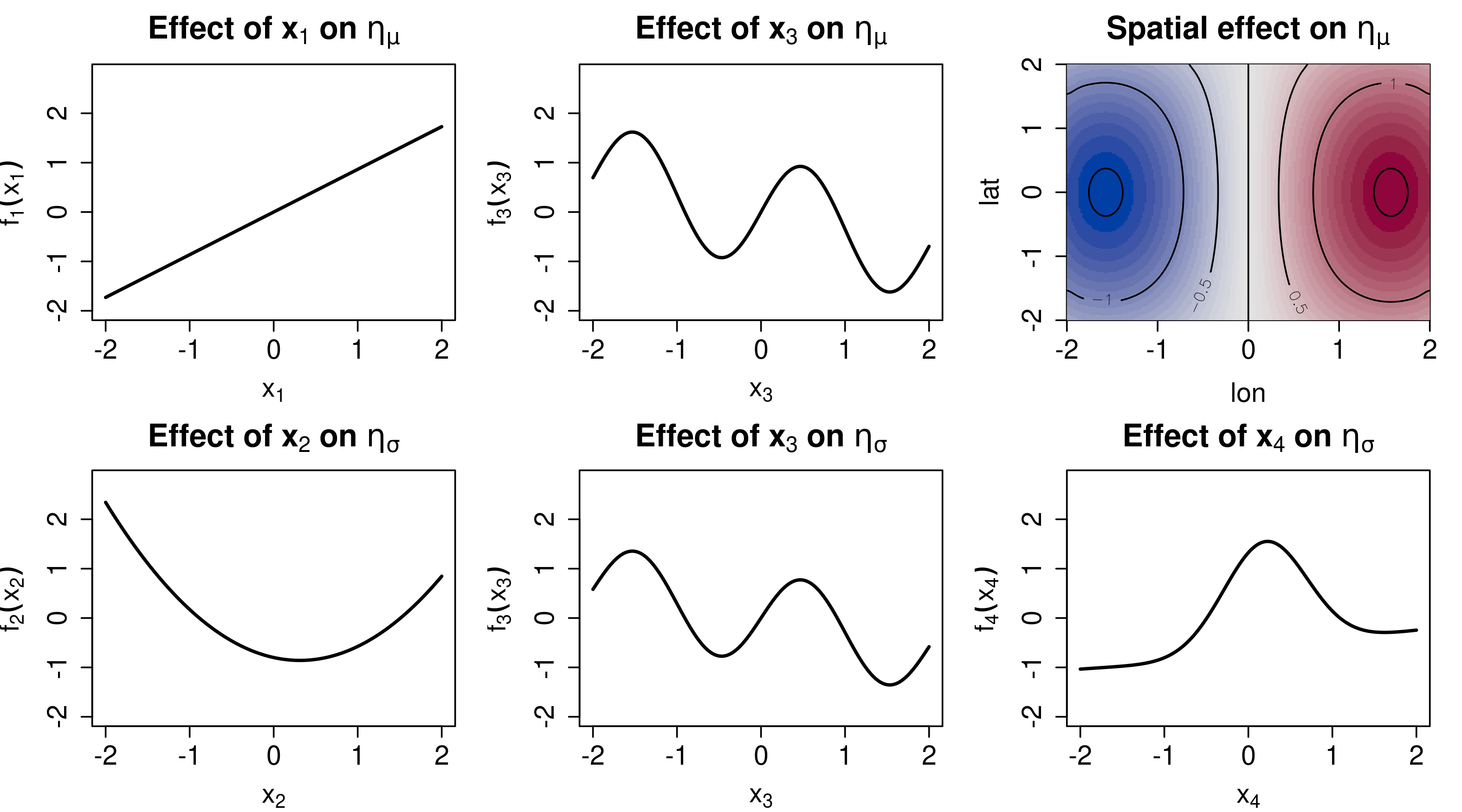}
\caption{\label{fig:NO_effects} Functions used in the simulation study.}
\end{figure}

\paragraph{Predictor Specifications}
We use the following  predictors $\eta_{\mu}$ and $\eta_{\sigma}$ for each distribution
\begin{eqnarray*}
\eta_{\mu} = \beta_{0\mu} + f_1(\texttt{x}_\texttt{1}) + f_3(\texttt{x}_\texttt{3}) +
  f_{2d}(\texttt{lon}, \texttt{lat}),\quad
\eta_{\sigma} = \beta_{0\sigma} + f_2(\texttt{x}_\texttt{2}) + f_3(\texttt{x}_\texttt{3}) + f_4(\texttt{x}_\texttt{4}),
\end{eqnarray*}
with model intercepts $\beta_{0\mu} = 0$, $\beta_{0\sigma} = 0$ for \code{NO},
$\beta_{0\mu} = 1$, $\beta_{0\sigma} = -1$ for \code{GA} and $\beta_{0\mu} = 1$,
$\beta_{0\sigma} = -1.5$ for \code{ZAP}; and 
\begin{eqnarray*}
f_1(x) &=& x\\
f_2(x) &=& x + ((2\cdot x - 2)^2) / 5.5\\
f_3(x) &=& -x + \pi \cdot \sin(\pi \cdot x)\\
f_4(x) &=& 0.5 \cdot x + 15 \cdot \frac{ \exp\left(- 2 \cdot (x - 0.2)^2\right) }{\sqrt{2\pi}} -
  \frac{\exp\left(-\frac{ (x +0.4)^2}{2}\right)}{\sqrt{2\pi}}\\
f_{2d}(z_1,z_2) &=& \sin(z_1) \cdot \cos(0.5 \cdot z_2).
\end{eqnarray*}
The simulated functions are shown in Figure \ref{fig:NO_effects}, these are centered around zero and 
scaled so that each effect has a similar range. 
The link functions for the
respective parameters are as follows: $\mu = \eta_{\mu}$,
$\log(\sigma^2) = \eta_{\sigma}$ for \code{NO}, $\log(\mu) = \eta_{\mu}$,
$\log(\sigma^2) = \eta_{\sigma}$ for \code{GA} and
$\log(\mu^2) = \eta_{\mu}$, $\log\left(\frac{\sigma}{1-\sigma}\right) = \eta_{\sigma}$
for \code{ZAP}. Finally, all covariates are drawn independently from uniform distributions
$\texttt{x}_\texttt{1}, \dots, \texttt{x}_\texttt{4}, \texttt{lon}, \texttt{lat} \sim \mathcal{U}(-2, 2)$.

\paragraph{Further settings}
\begin{itemize}
\item To investigate performance for small and large data settings alike, we simulate $n=$500, 1000, 10000 and 50000 number of observations.
\item To challenge variable selection, an additional number of noise variables (denoted with
  $\texttt{nnoise} = 0, 10, 20$ in the following)
  is considered. Each predictor is modeled including all
  available covariates. Accordingly, for each predictor three true covariates and
  \code{nnoise} non-relevant covariates are included. Note that variables \code{lon}, \code{lat} 
  are counted as one covariate.
\item In the first case the covariates are uncorrelated ($\rho=0$), and in the second case correlation
  is introduced by the Cholesky factorization $\mathbf{L}\mathbf{L}^\top = \boldsymbol{\Sigma}$ of the covariance matrix
  $$
  \boldsymbol{\Sigma} = \begin{pmatrix}
    1 & \rho & \rho^2 & \cdots & \rho^{l-1} \\
    \rho & 1 & \rho & \cdots & \rho^{l - 2} \\
    \vdots & \vdots & \vdots & \ddots & \vdots \\
    \rho^{l - 1} & \rho^{l - 2} & \rho^{l - 3} & \cdots & 1
  \end{pmatrix},
  $$
  where $l$ is the number of covariates, and correlated covariates are thus generated with
  $\mathbf{X}_{\text{corr}} = \mathbf{X}\mathbf{L}^\top$ with $\rho = 0.7$.
\item Each scenario is replicated 100 times.
\end{itemize}
\paragraph{Measures of Performance}  To evaluate the performance of
  the four  algorithms, the MSE of the predictors, the MSE of the effects, the continuous ranked probability score 
  \citep[CRPS;][]{sdr:Gneiting+Raftery:2007} and the number of falsely selected variables in each
  predictor (false positives) are calculated based on an \textit{out-of-sample} validation
  data-set with $10000$ observations. The validation data-set is fixed throughout each response distribution and each
  $\rho \in \lbrace 0, 0.7 \rbrace$. We define the MSE of the predictor as the mean of the squared differences of the estimated additive
  predictors and the true additive predictors
  ($\text{MSE}_k = \frac{1}{n} \sum_i (\hat{\eta}_{i,k} -\eta_{i,k})^2$, for $k = \mu,\sigma$). For the MSE of the effects we use a similar notation, namely the mean of the squared differences of the true effects and the estimated effects ($\text{MSE}_{f_{k,j}} = \frac{1}{n} \sum_i (\hat{f}_{i,k,j} -f_{i,k,j})^2$, for $k = \mu,\sigma$ and $j = 1,2,3,4,2d$).
The false positive rate is defined as the number of
non-informative covariates which have a sufficiently large estimated effect $f$, i.e.
$max(f) - min(f) >$ \code{threshold}, with \code{threshold = 0.1}.

\paragraph{Computational Details}
The simulation was run on the HPC infrastructure LEO4 of the University of Innsbruck. This
HPC infrastructure runs on a Linux system (CentOS 7), and 50 computing nodes with Intel Xeon
(Broadwell/Skylake) processors with up to 3000 gigabyte (GB) available memory.
Depending on the setting, the memory requirements are between 5 and 50 GB per replication.

\subsection{Results}

\begin{figure}[ht!]
\centering
\includegraphics[width=0.75\textwidth]{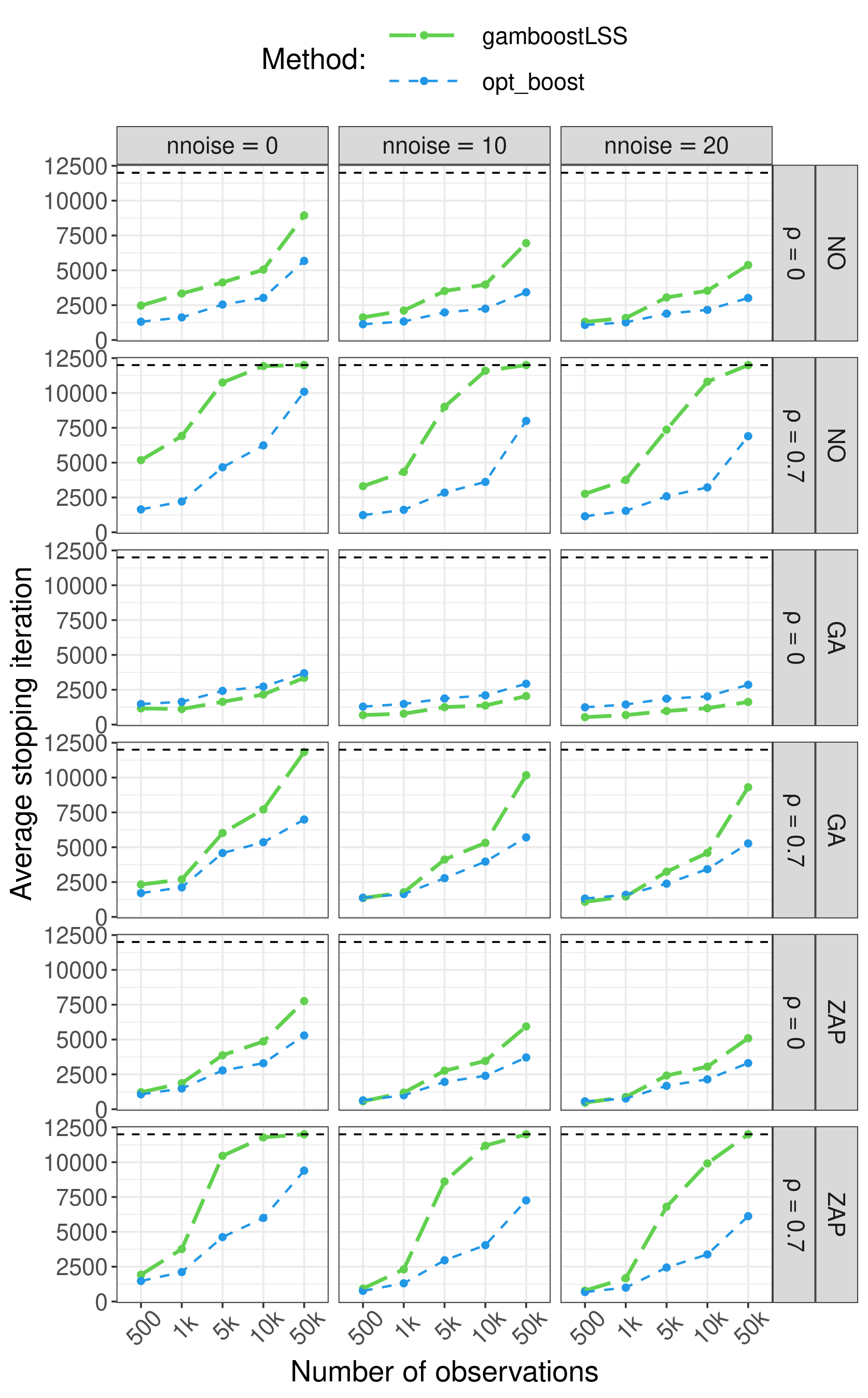}
\caption{\label{fig:resmstop} Simulation study. Average $\mathit{mstop}$ of both boosting variants.
  Correlation $\rho = 0.7$ leads to higher $\mathit{mstop}$.
  For higher $n$ and correlation the \code{gamboostLSS} variant hits
  the upper limit (horizontal dashed line at $12000$) of possible stopping iterations.}
\end{figure}

\begin{figure}[ht!]
\centering
\includegraphics[width=0.75\textwidth]{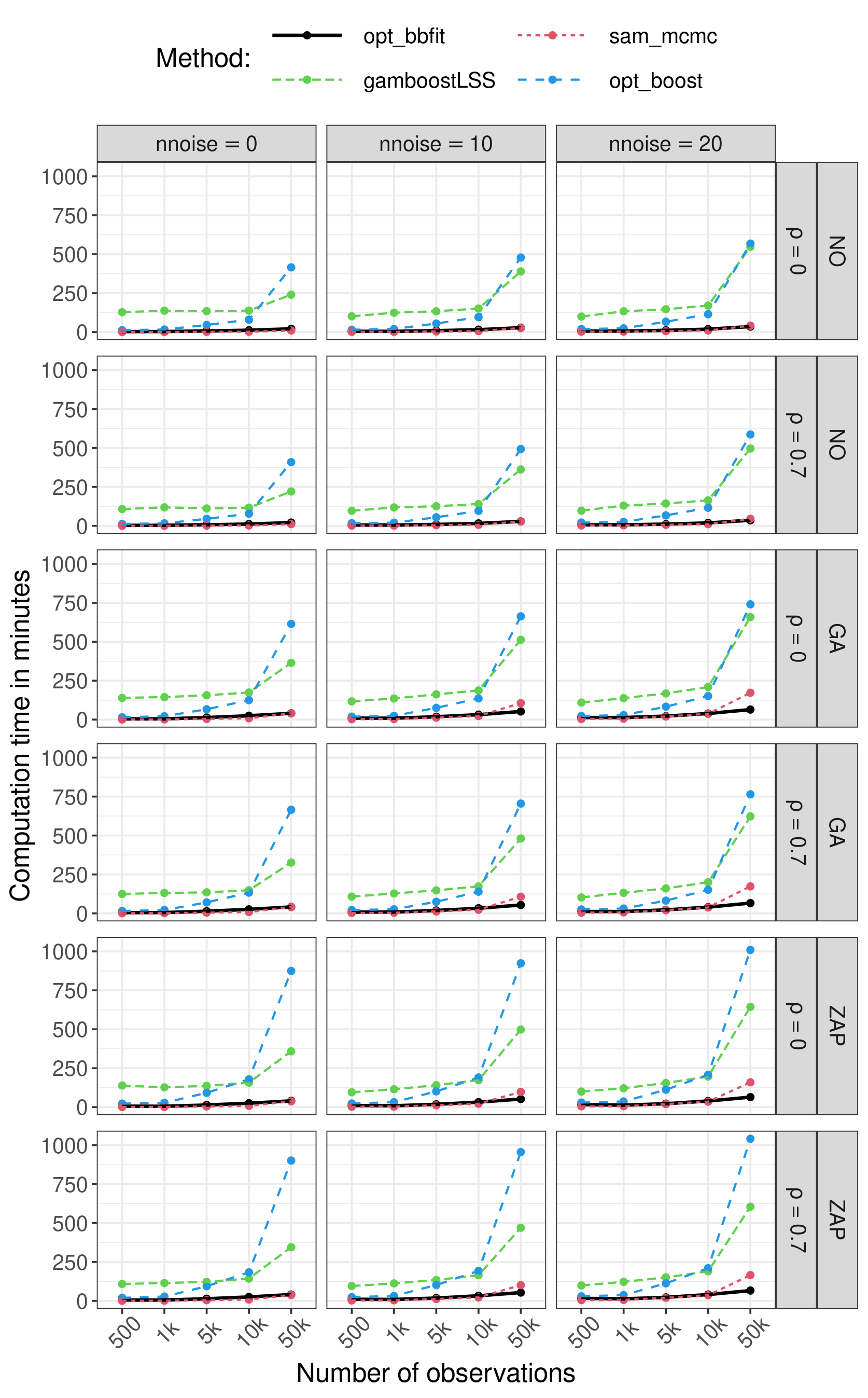}
\caption{\label{fig:reselapsed} Simulation study. Computation times for all distributions,
  noise variables and correlation settings. The most computational intensive settings\textemdash the two
  boosting variants\textemdash depending on the number of observations require up to approximately
  18 hours to compute. In the same settings \code{opt\_bbfit} needs around $1$ hour,
  while the MCMC method runs $3$ hours.}
\end{figure}

\paragraph{Stopping and Computing Times}
Due to high computing times, we set the number of maximum iterations to identify the optimal stopping iteration $\mathit{mstop}$ for the two boosting algorithms to 12000.
Figure \ref{fig:resmstop} shows the average $\mathit{mstop}$ of all settings. For both boosting methods the average $\mathit{mstop}$ increases with
the sample size and with larger correlations between covariates. In all but the non-correlated \code{GA}
settings, \code{opt\_boost} has a lower average of $\mathit{mstop}$ than  \code{gamboostLSS}. With increasing $n$ and $\rho = 0.7$, the average $\mathit{mstop}$ is $12000$
for \code{gamboostLSS}, indicating that more iterations are needed. Figure \ref{fig:reselapsed}
shows the elapsed time in minutes for each setting and method.
The \code{gamboostLSS} boosting
method needs around 600 to 1100 minutes to compute a single simulation run when $n=50000$ (similar to \code{opt_boost} which also needs several hours). Thus we deem increasing the maximum of available iterations as infeasible.
In contrast, the batchwise backfitting method needs only 30 to 60 minutes for these settings.

\begin{figure}[t!]
\centering
\includegraphics[width=0.75\textwidth]{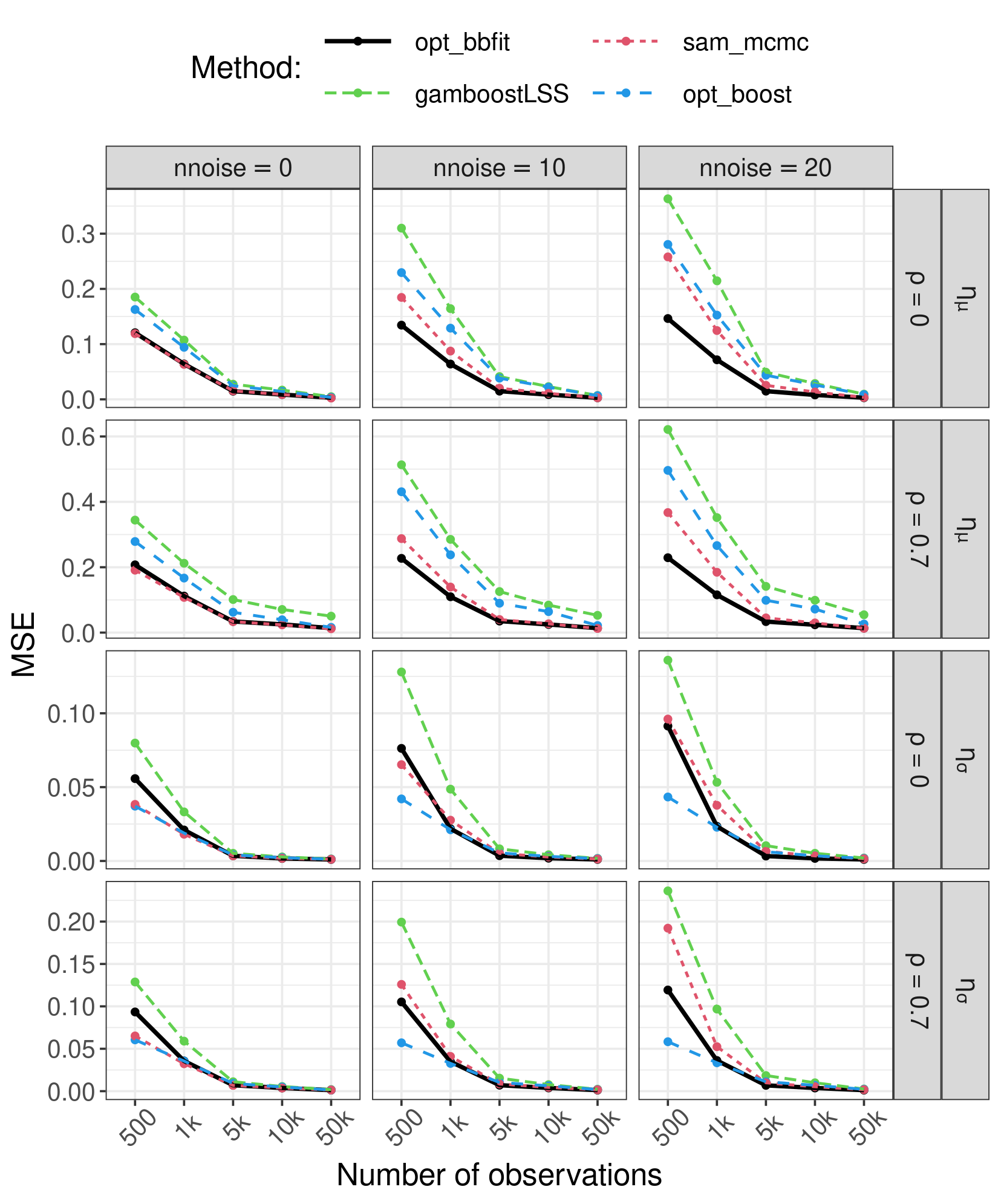}
\caption{\label{fig:resNO} Simulation study. Average MSE with the \code{NO} distribution for
  predictor $\eta_{\mu}$ and $\eta_{\sigma}$ for different number of observations,
  correlation and noise variable settings.}
\end{figure}

\begin{figure}[ht!]
\centering
\includegraphics[width=0.71\textwidth]{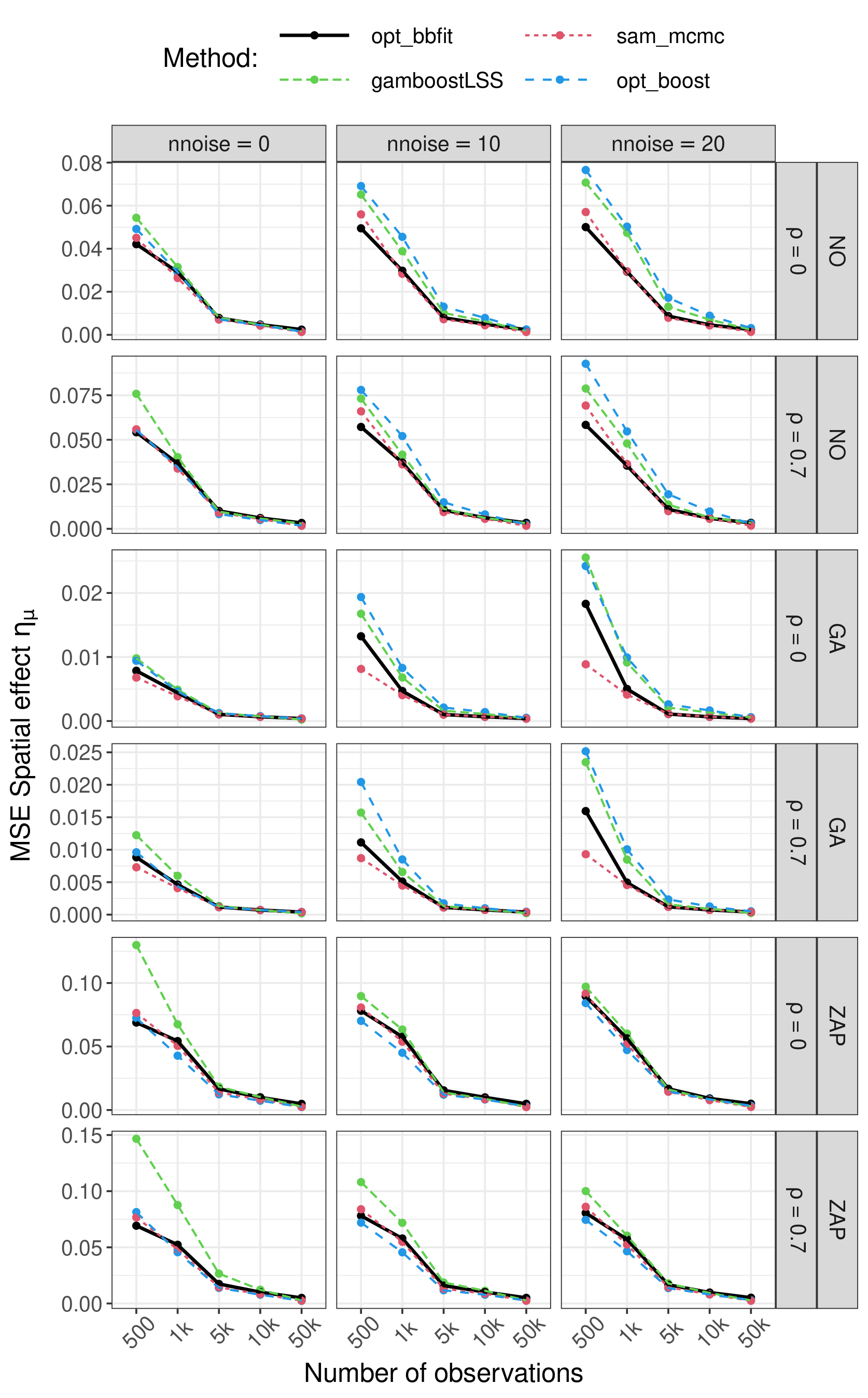}
\caption{\label{fig:f2d}  Simulation study. Average MSE of spatial effect of $\eta_{\mu}$ for all distributions,
  different number of observations, correlation and noise variables.}
\end{figure}

\begin{figure}[ht!]
\centering
\includegraphics[width=0.71\textwidth]{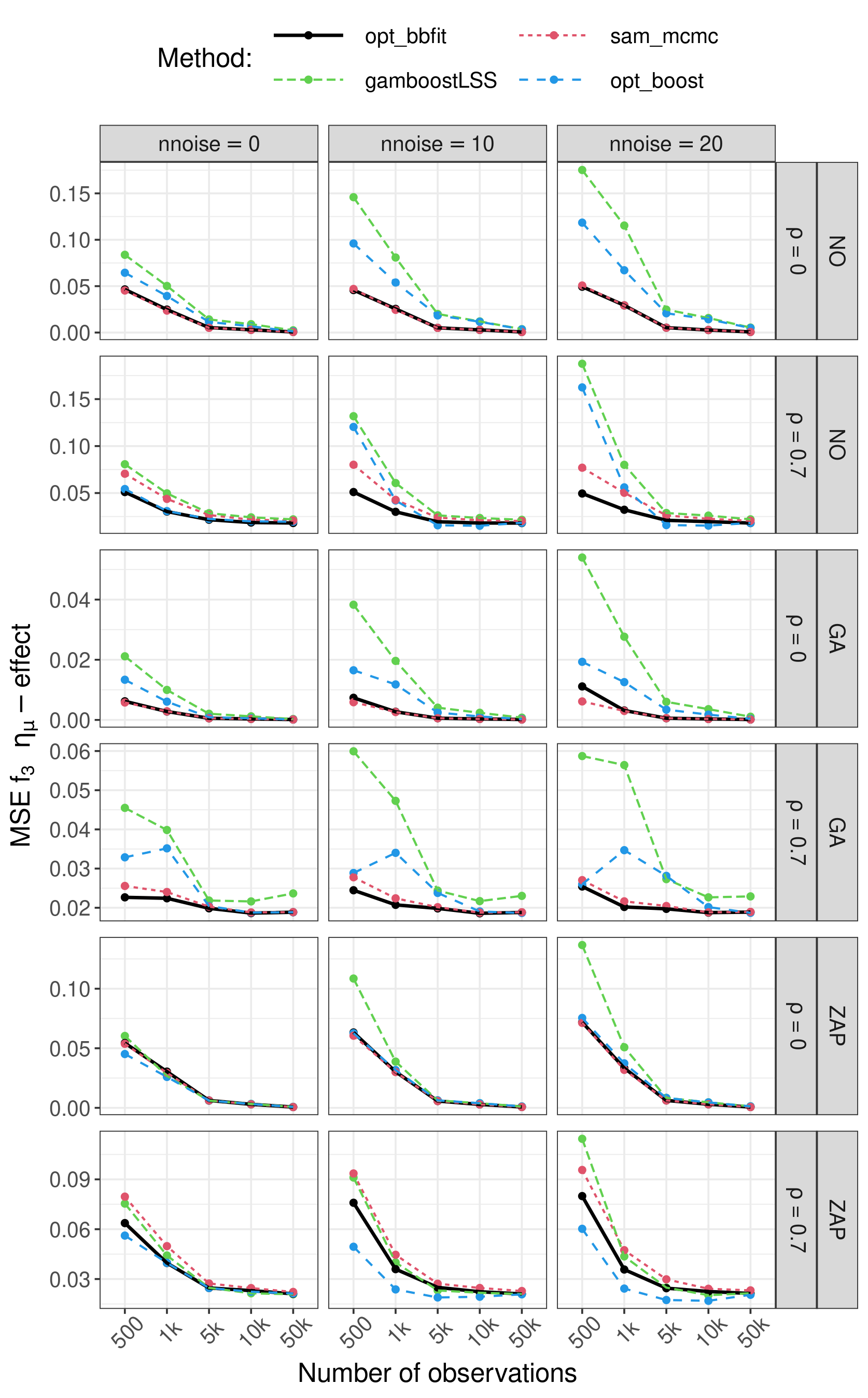}
\caption{\label{fig:f3}  Simulation study. Average MSE of $f_3$ effect of $\eta_{\mu}$ for all distributions,
  different number of observations, correlation and noise variables.}
\end{figure}

\paragraph{MSE}
A comparison of the four methods in terms of MSE is made in Figure \ref{fig:resNO} for all
\code{NO}-settings. The MSE decreases sharply with increasing $n$, except for \code{gamboostLSS} in $\eta_\mu$ with $\rho = 0.7$,
this method has a much higher MSE here. This is due to the limited number of stopping iterations available for
\code{gamboostLSS} (note again, that the stopping iteration is set very large to $12000$).
The \pkg{bamlss} methods perform better in the small $n$ settings than the \code{gamboostLSS} method.
Remarkably, the method \code{opt_bbfit} has the smallest MSE for $\eta_{\mu}$ when it includes noise variables, and has basically
the same performance as \code{sam\_mcmc} without noise variables. Only in $\eta_{\sigma}$ and $n = 500$ is \code{opt\_bbfit}
second or third best in each case, though it always performs better than \code{gamboostLSS}.
For larger $n$, the methods are very similar in terms of MSE. The results for the \code{GA} and \code{ZAP}
distribution are qualitatively very similar and they can be found in the
Appendix~(Figures \ref{fig:resGA} and \ref{fig:resZAP}).
For the individual effects biases, i.e. the MSE of the effects, we refer to the Figures  \ref{fig:f2d} and \ref{fig:f3} and the Appendix~(Figures \ref{fig:f1}, \ref{fig:f2s}, \ref{fig:f3s} and \ref{fig:f4s}). The overall result is that the \code{opt\_bbfit} is very competitive in terms of MSE of the effects,  in a lot of settings it performs better than the boosting methods, although the MSE of the effects converges for all methods and effects close to zero with high enough numbers of observations.  

\paragraph{Predictive Accuracy}
\begin{figure}[ht!]
\centering
\includegraphics[width=0.71\textwidth]{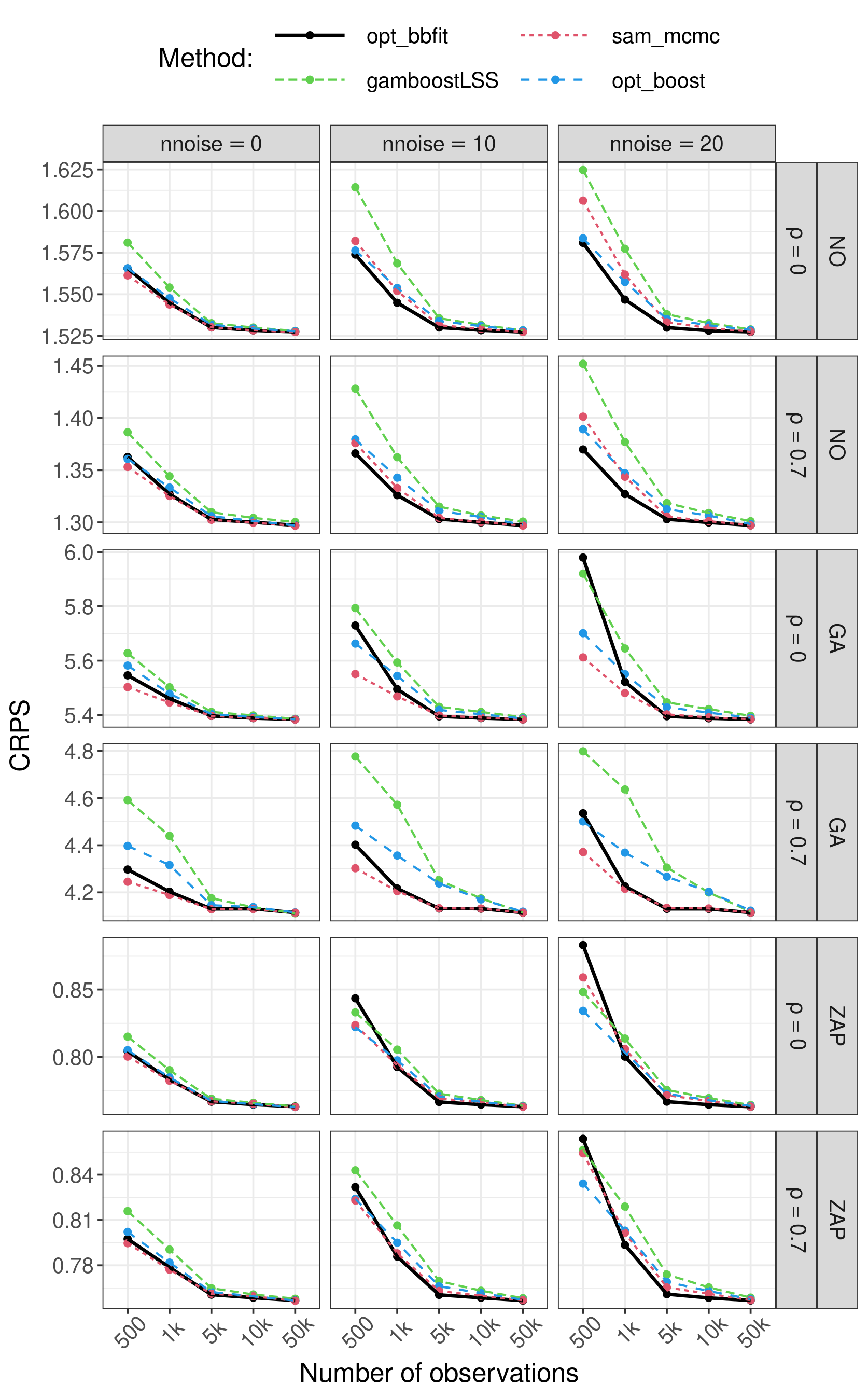}
\caption{\label{fig:rescrps}  Simulation study. Average CRPS for all distributions,
  different number of observations, correlation and noise variables.}
\end{figure}
Figure \ref{fig:rescrps} shows the CRPS of the three different distributions with smaller values indicating
higher predictive accuracy. The results are very similar to the results of the MSE. It is again noteworthy
that \code{opt\_bbfit} has the best performance when noise variables are included in the  \code{NO} settings 
for all $n$, and in the other settings, when $n\geq 1000$, the performance is almost identical to \code{sam\_mcmc} 
and typically better than the boosting methods. For larger data settings the methods are very similar in terms 
of CRPS.

\paragraph{Variable Selection}
\begin{figure}[t!]
\centering
\includegraphics[width=0.9\textwidth]{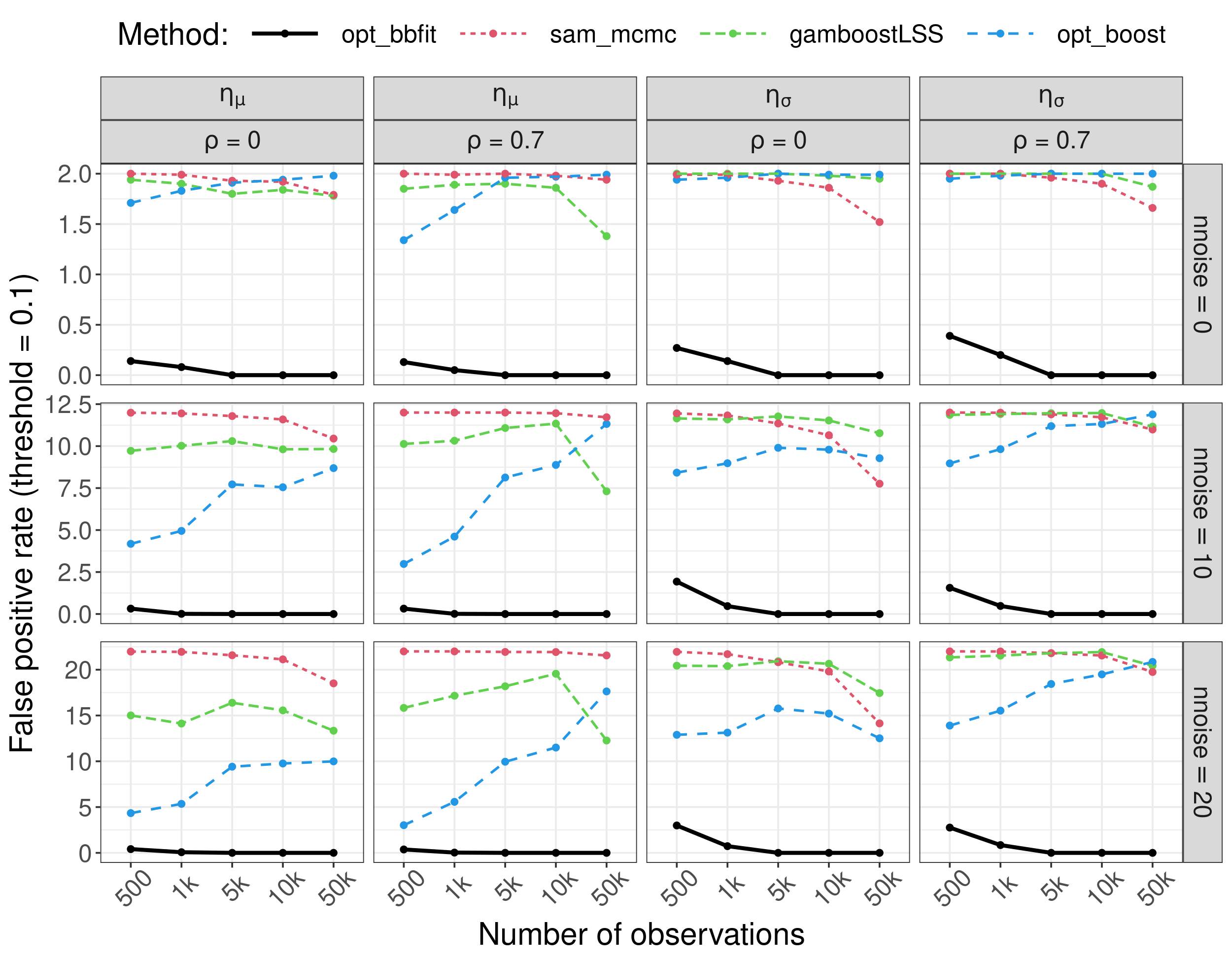}
\caption{\label{fig:resNO_fp} Simulation study. Average false positive rate with the \code{NO} distribution for
  predictor $\eta_{\mu}$ and $\eta_{\sigma}$ for different number of observations,
  correlation and noise variable settings.}
\end{figure}

The average false positive rates with the \code{NO} distriubtion are displayed in Figure \ref{fig:resNO_fp}. The false
positive rate is defined as the number of non-informative covariates which have a sufficiently large estimated
effect $f$, i.e. $max(f) - min(f) > \texttt{threshold} = 0.1$.
The proposed batchwise backfitting method \code{opt\_bbfit} outperforms every other method in terms of
false positive rates. In all, except two \code{GA} distribution settings, $n \geq 5000$ observations are
always sufficient to exclude all non-informative covariates with the novel approach. In all settings, the
true positive rates are 1 for all methods (not shown). We also evaluate the false positive rate with a whole
range of thresholds starting very restrictive from $0.0001$ up to $0.3$ for a \code{NO}-setting with different
numbers of observations ($n = 250, 500, 5000, 10000$) and find that except in
the $n = 250$ case the \code{opt\_bbfit} is performing best
(see Appendix~Figure~\ref{fig:resthres}). The results for the \code{GA} and \code{ZAP}
distribution are qualitatively the same and can be found in the Appendix
(Figures \ref{fig:resGA_fp} and \ref{fig:resZAP_fp}).

\paragraph{Summary}
Our batchwise backfitting algorithm has basically the same perfomance on small datasets ($n \leq 1000$)
in terms of MSE and CRPS compared to the other three methods used in this study, and on medium and
large datasets ($n \geq 5000$) it is almost consistently the best method.
Compared to boosting, where computationally intensive CV (or similar) is needed to determine
$\mathit{mstop}$  batchwise backfitting does not require such additional time-consuming tuning.
This makes our algorithm particularly convenient when applied on very large data sets. Our novel
method is also considerably faster than both boosting variants (even with a determined $\mathit{mstop}$.
The speed advantage ranges from around five times faster for $n = 500$ up to $15$ to $20$ times
faster in the large data setting with $n = 50000$ observations. Compared with the \pkg{bamlss}
MCMC implementation, the speed advantage is evident from $n = 50000$ settings,
and we expect it to increase dramatically in even larger data settings. The
false positive rates of our batchwise backfitting method are excellent in all settings which makes
this method also an ideal option for variable selection. Please note once more, that due to
computational costs of benchmark methods the maximum number of observations was $50000$ only.
However, in the next Section~\ref{sec:lightapp} we show a model estimated with
$\approx 9.1$ million observations and in the Appendix~\ref{appendix:estimation} we exemplify
that our batchwise backfitting method can easily handle up to $10^7$ and more observations.

\section{Application: Lightning Count Model} \label{sec:lightapp}

Lightning is a major source of atmospheric nitrogen oxides \citep{schumann2007}
which is an important greenhouse gas \citep[Figure SPM.2 in][]{ipcc2021}. Thus, lightning affects
the climate. At the same time lightning is affected by climate change. This effect
is subject to scientific debate \citep{murray2018}. As lightning processes cannot be
resolved by numeric models of the atmosphere, this debate is mainly
based on proxies of lightning that have a simple formulation and often
consider only a particular aspect of the physical processes involved in lightning.
Such simple formulations might be the cloud top height \citep{price1992},
iceflux in the mid atmosphere \citep{finney2014} or wind shear \citep{taszarek2021},
among others.

As a reaction, scholars proposed to analyse lightning using  machine learning (ML) approaches
that incorporates numerous physical processes \citep[e.g.,][]{ukkonen2019, Simon+Mayr+Morgenstern+Umlauf+Zeileis:2021}.
These  are capable to process large amounts of data and
identify most relevant variables from a pool of inputs, but focus on describing the
occurrence of lightning via binary classification and \emph{not} on the number of
lightning counts which would be crucial to investigate the important quantity of
\emph{flash rates} \citep{cecil2014}.

The batchwise backfitting method proposed in this manuscript allows, for the first time, the 
estimation of a high-dimensional, fully probabilistic count data model, including variable 
selection, using a very large data set.

\paragraph{Data}
We use high-resolution data from the Austrian Lightning Detection and
Information System \citep[ALDIS, ][]{schulz2005} and explain the lightning counts with
reanalysis data from ERA5, the fifth generation of ECMWF (European Centre for Medium-Range Weather
Forecasts) atmospheric reanalyses of global climate \citep{era5a, era5b}.
ERA5 provides globally complete and consistent pseudo-observations of the atmosphere using
the laws of physics. The horizontal resolution is approx.\ $32~km$, while the temporal resolution
is hourly and covers the years from 1950 to present. The model is not only interesting for
a more comprehensive description of lightning,
but also for a full reanalysis to study climate trends in lightning
\citep{Simon+Mayr+Morgenstern+Umlauf+Zeileis:2021}, because homogeneous lightning observations
from ALDIS are only available for the period in the order of a decade, here 2010--2019.

\begin{figure}[t!]
\centering
\includegraphics[width=1\textwidth]{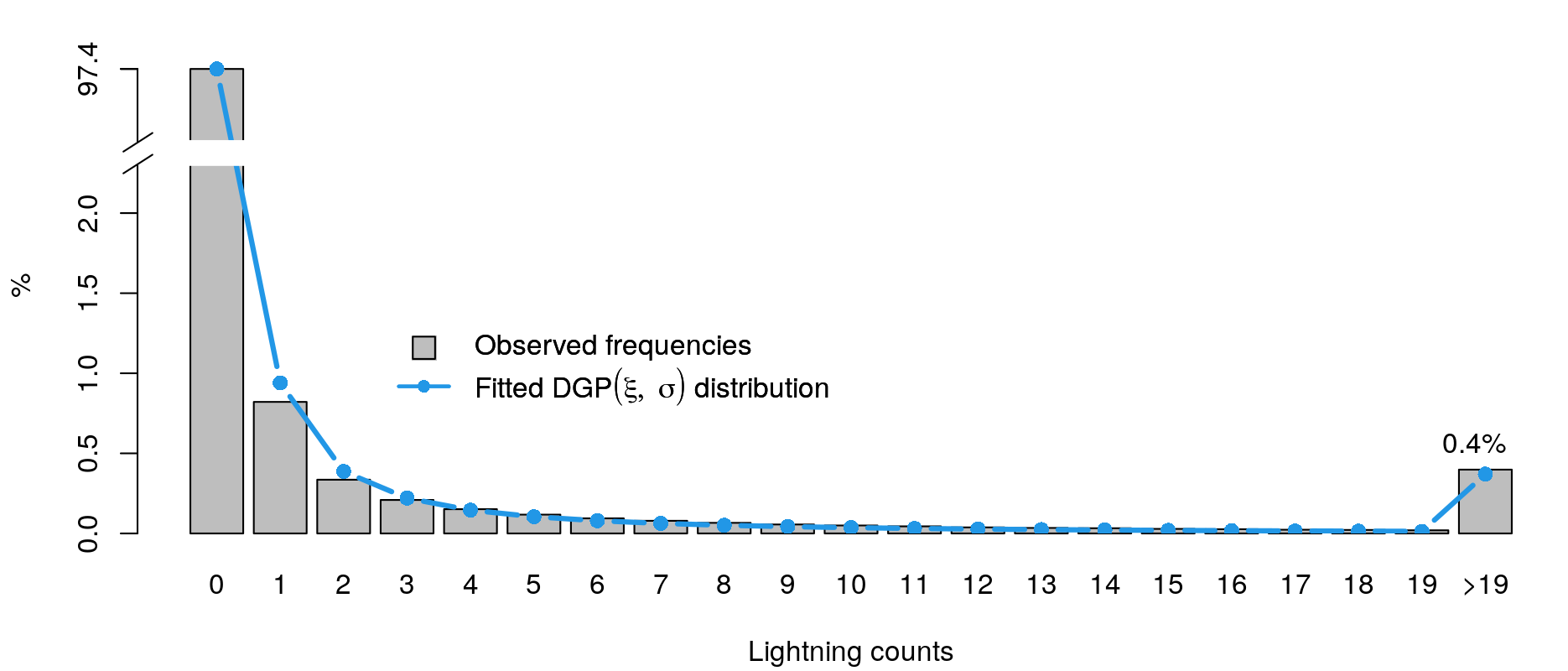}
\caption{\label{fig:lightning_counts} Distribution of hourly lightning counts, gray bars,
  along with estimated frequencies using a discretized generalized Pareto distribution,
  $\texttt{DGP}(\xi, \sigma)$, blue dots and lines. Note the y-axis is broken because of the
  large number of zeros in the data, $97.4\%$.}
\end{figure}

We develop a model for the complete lightning count distribution
using our proposed batchwise backfitting algorithm from Section~\ref{sec:bbfit}.
Therefore, we aggregate the hourly lightning counts to the ERA5 grid cells, resulting in a final
data set of $\approx 9.1$ million observations. To estimate a well-calibrated model,
we preselect 76~ERA5 covariates that are potentially good candidates for lightning and convective processes,
such as convective available potential energy (\code{cape}), convective precipitation (\code{cp}),
cloud top height (\code{cth}), specific cloud snow water content between $-20^{\circ}C$ and $-40^{\circ}C$ (\code{cswc2040}),
among others \citep[for a detailed description of variables see][]{morgenstern2022supp}. Since the distributional model
is quite complex and very many covariates also have strong skewness, these are
standardized before estimation using the empirical cumulative distribution function estimated with
the training data, so that all covariates are in the value range $[0, 1]$ and thus numerical
problems can be avoided
To examine the final model performance, we split the data into a training and a test data set with
$\approx 8.2$ million (2010--2018) and $\approx 0.9$ million (2019) observations, respectively.
The distribution of hourly lightning counts is shown in Figure~\ref{fig:lightning_counts} and
indicates that the data contain a very large number of zero counts.

\begin{figure}[t!]
\centering
\includegraphics[width=1\textwidth]{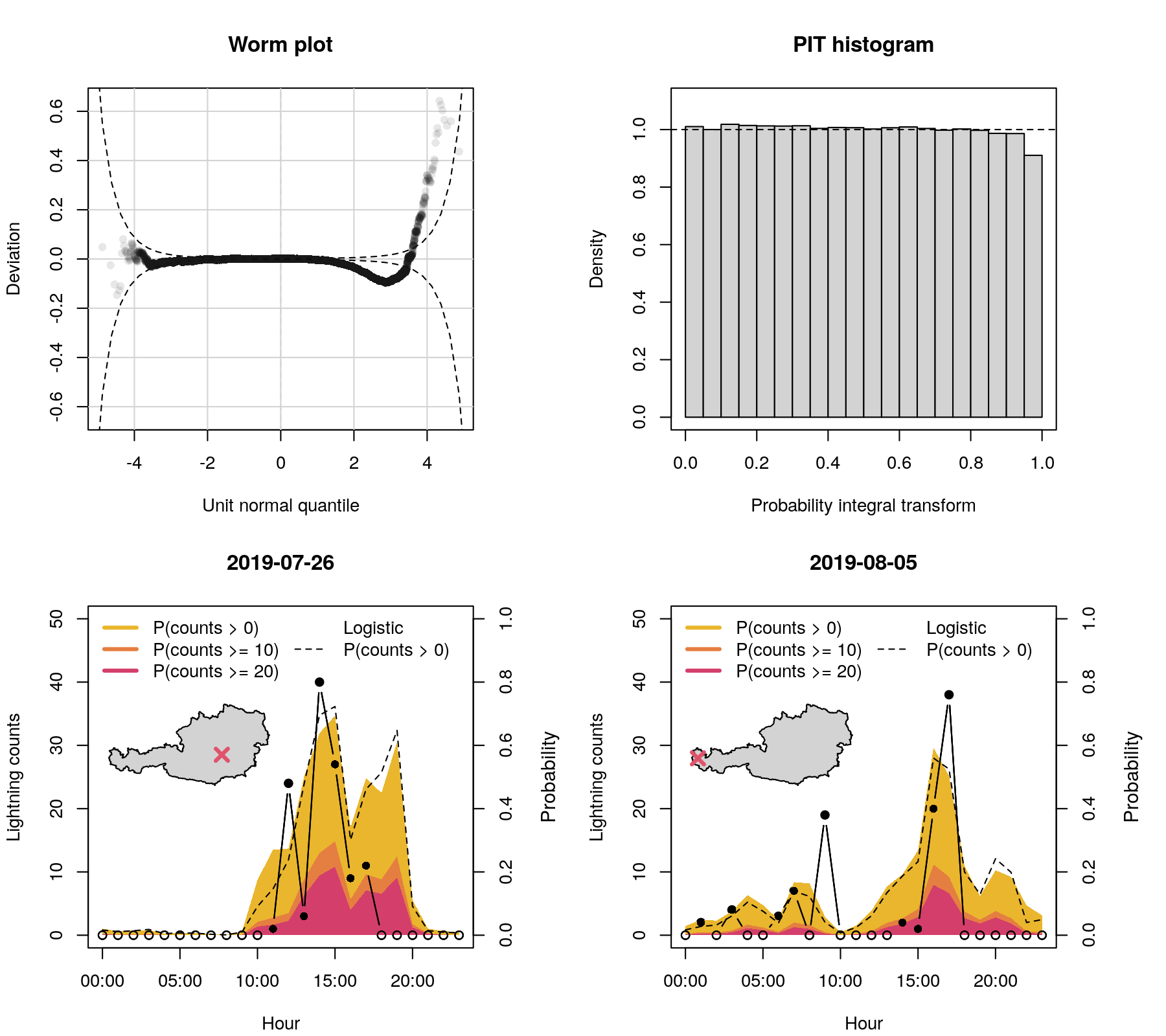}
\caption{\label{fig:lightning_prediction}
  \code{DGP} lightning model. The top left panel shows a worm plot of the
  out-of-sample randomized quantile residuals. The top right panel the corresponding
  probability integral transform (PIT) histogram. Out-of-sample predictions from the \code{DGP}
  lightning model for two locations and dates are shown in the second row. Observed lightning
  counts are represented by the black dot line, predicted probabilities are shown in the
  background in heat colors. Predictions for lighting yes/no from a logistic model are shown by
  the black dashed lines. Prediction location is shown by the red cross in the map of Austria.}
\end{figure}

\paragraph{Model Specification}
For this reason, in a distributional model for the number of lightnings, the extreme frequency of
zeros must be considered. We found that the discretized version of the generalized
Pareto distribution, \code{DGP}($\xi$, $\sigma$), provides promising results (see the first row of  
Figure~\ref{fig:lightning_prediction} and the next paragraph).
For details on construction of the \code{DGP}, and
discrete distributions in general we refer to \citet{Subrata:2015, KRISHNA2009177}.

As a first overall check, we fitted an intercept only model using the batchwise backfitting
algorithm to assess the goodness of fit of the unconditional distributional model with \code{DGP}.
The model estimates the two parameters with a batchsize of 50000 and 1000 batches within
about 5 minutes on a conventional laptop with Intel(R) Core(TM) i7-8550U CPU \@ 1.80GHz processor.
This fitted \code{DGP} density is shown in Figure~\ref{fig:lightning_counts} by the blue dots and lines and indicates
that the model follows the observed relative frequencies well. However, some probabilities are overestimated,
e.g., for one and two lightnings, which is due to the fact that the model does not yet include covariates.

Thus, we consider the following prediction model
$\texttt{counts} \sim \texttt{DGP}(\log(\xi) = \eta_{\xi}, \log(\sigma) = \eta_{\sigma})$ and
additive predictors
$$
\eta_k = f_{1k}(\texttt{doy}) + f_ {2k}(\texttt{hour}) + f_{3k}(\texttt{lon}, \texttt{lat}) +
  f_{4k}(\texttt{cape}) + \ldots + f_{80k}(\texttt{mcc}),
$$
where covariate \code{doy} is the day of the year, \code{hour} the hour of the day and model
term $f_{3k}(\texttt{lon}, \texttt{lat})$ specifies a spatial effect of longitude and
latitude coordiantes. Model terms $f_{4k}( \cdot ), \ldots, f_{80k}( \cdot )$ represent the
effects of further ERA5 covariates, such as convective available potential energy (\code{cape}) or
the medium cloud cover (\code{mcc}). For the scope of simplicity, we do not elaborate on these 
covariates in this manuscript; we refer the reader to \citet{era5a, era5b} for details.

\paragraph{Model Fitting}
For fitting this model we proceed as follows. 
We use the boosting variant of the batchwise backfitting algorithm to select
the most suitable of the 80 model terms. We use the AIC for selecting suitable covariates with
200 batches of size 50000. The estimation time is about 12 hours, which is not very long
considering the huge data set and the very large number of covariates.
\begin{figure}[t!]
\centering
\includegraphics[width=1\textwidth]{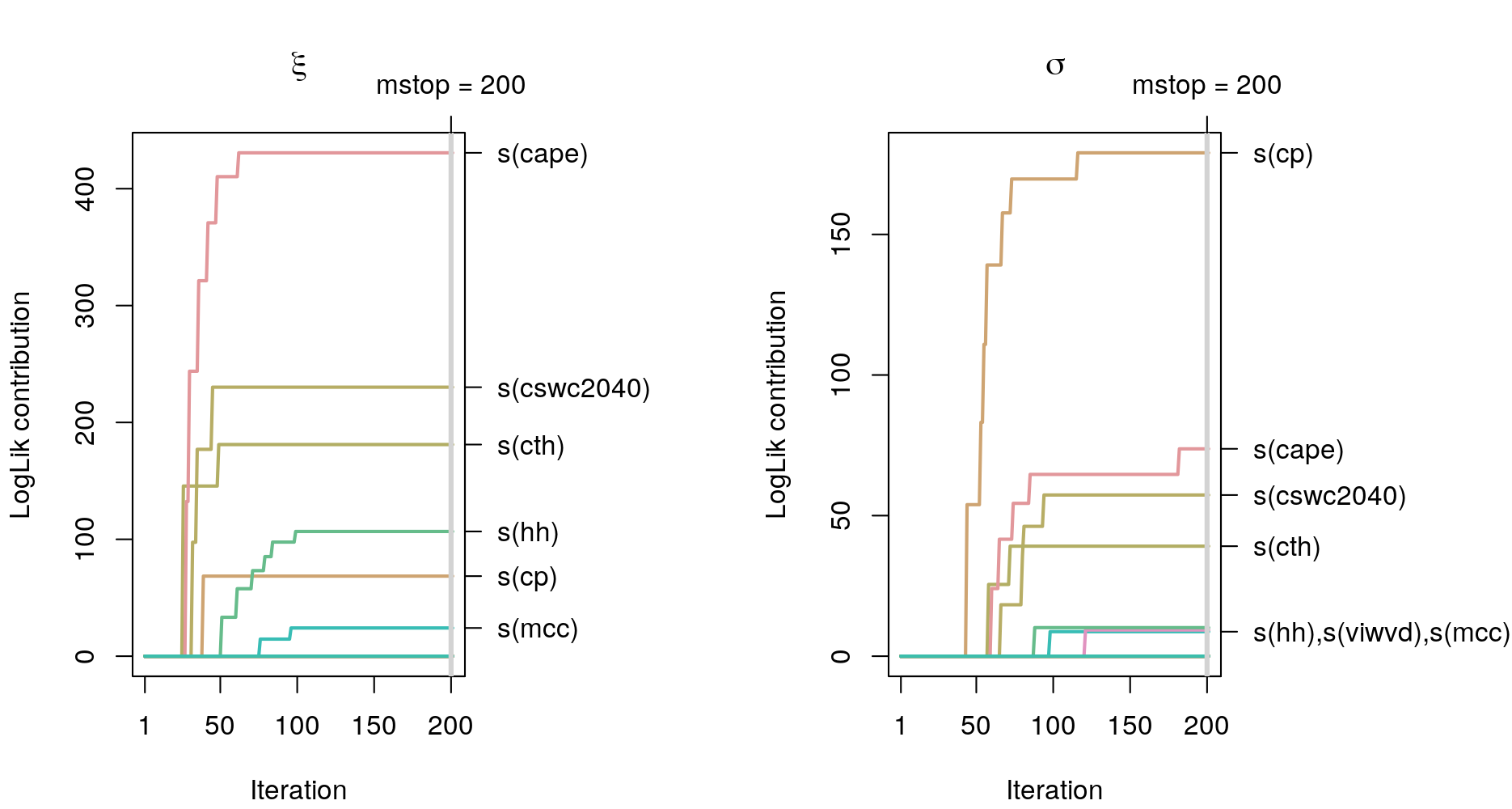}
\caption{\label{fig:lightning_contribplot} Batchwise backfitting log-likelihood
  contributions for parameters $\xi$ and $\sigma$ of selected covariates.}
\end{figure}
\begin{figure}[t!]
\centering
\includegraphics[width=1\textwidth]{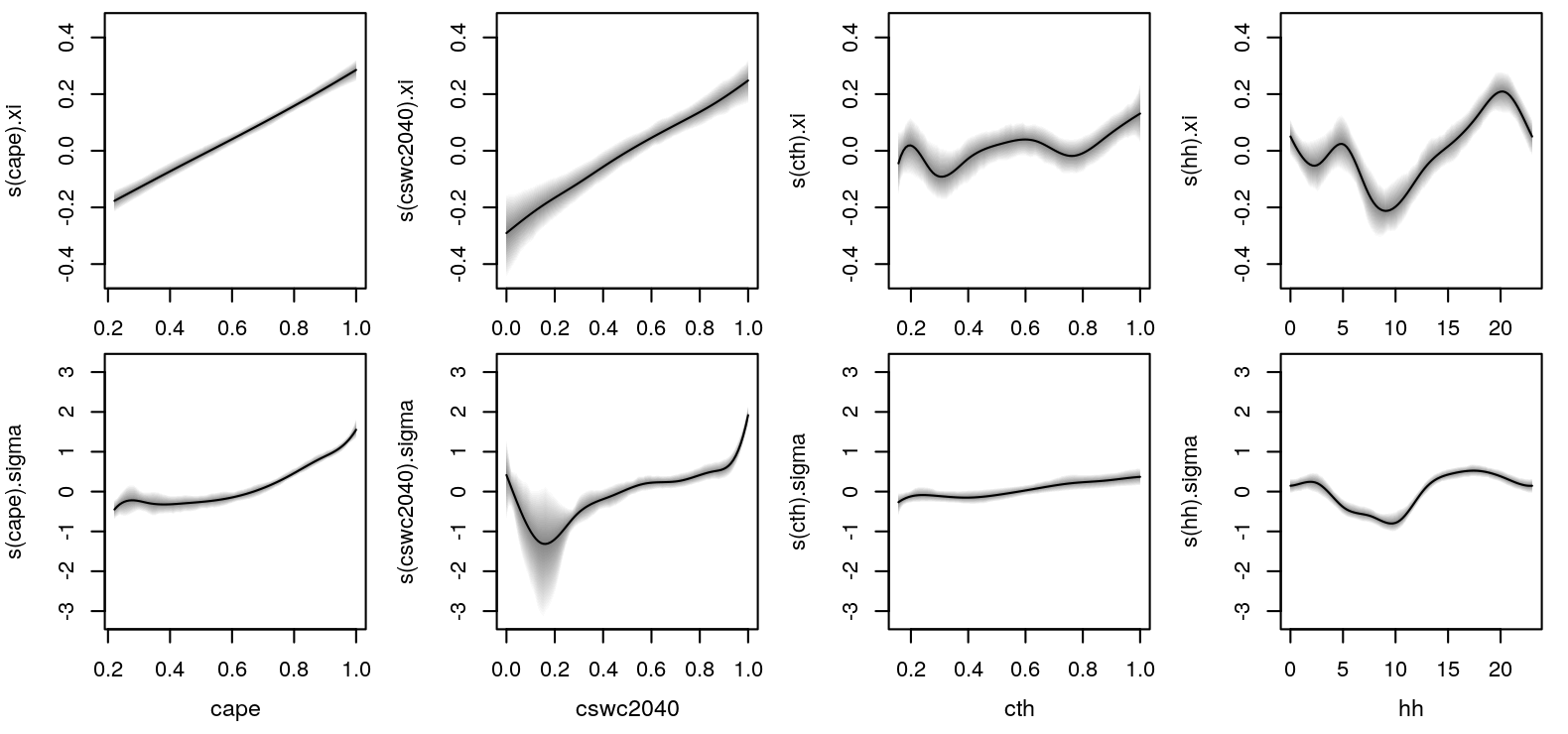}
\caption{\label{fig:lightning_effects}
  \code{DGP} lightning model. Selection of estimated smooth effects of the
  final model for parameter $\xi$, top row, and parameter $\sigma$, bottom row.
  The gray shaded areas show the variation of estimates in the resampling variant
  of the batchwise backfitting algorithm.}
\end{figure}
In Figure~\ref{fig:lightning_contribplot} the log-likelihood
contributions for the selected covariates are shown indicating that the algorithm
converged running 200 iterations/batches. Note that using the AIC in the
batchwise backfitting algorithm results in a rather sparse
model that selects only the most relevant variables, and that the final selected covariates are
consistent with the study of \citet{Simon+Mayr+Morgenstern+Umlauf+Zeileis:2021} which use
binary classification only. This fact is noteworthy because variable selection is done for the
entire data set and in one run of the batchwise backfitting algorithm, as opposed, e.g., to
the costly CV commonly used in boosting such models.
In a second step, the model is refitted with the selected variables using the resampling variant of
the batchwise backfitting algorithm to improve the predictive performance.
We again use 200 batches of size 50000, not using the first 100 iterations as burn-in.
The estimation time is approximately 35 minutes using the training data with $\approx 8.2$
million observations. The reason why the resampling variant is so much faster is mainly due
the very sparse model and the use of slice sampling of the smoothing variances under the
``out-of-sample'' AIC (see Section~\ref{sec:hyper}).

\paragraph{Results}
The estimated smooth effects of the final model are shown in Figure~\ref{fig:lightning_effects}. The 
effects show that some of the covariates could be modeled by linear functions, such as the effects 
for the variables \code{cape} and \code{cswc2040} for the parameter $\xi$. Instead, others, such as 
the effect for \code{hh} for both $\xi$ and $\sigma$, are nonlinear. The estimated effects appear 
plausible, e.g., an increase in \code{cape} for the location parameter $\xi$ increases the number of 
lightning counts, shifting the probability mass of the \code{DGP} distribution to larger counts. 
Similarly, for the scale parameter $\sigma$, increasing \code{cape} also results in a 
shift in the probability mass towards larger lightning counts. The effects for \code{hh} also 
show that higher counts can be expected in the afternoon, when the ground air temperature reaches 
its maximum.

In the first row of Figure~\ref{fig:lightning_prediction}, a worm plot is shown along with a
probability integral transform (PIT) histogram of the quantile residuals calculated using the test
data (year 2019). Both plots show that the model is quite well calibrated, only for
the very large count observations (about 2\% in the worm plot) the model does not seem
to be optimally balanced. The reason for this is certainly the extremely low number of
cases for large lightning counts, these are simply extremely difficult to model as a result.
In the second row of Figure~\ref{fig:lightning_prediction} we show 
the prediction from the \code{DGP} model together with the observed lightning counts for two days 
and locations in the test data set. The predictions show well that the model is indeed able to 
reflect the observed lightning activity. In addition, we also show in the plot for comparison the 
prediction from a logistic model for lightning yes/no (using the same selected covariates), marked 
by the black dashed line. It can be seen that the \code{DGP} model and the binomial model give 
basically the same point predictions, however the \code{DGP} model is much more informative as it 
allows to to derive prediction probabilities at different thresholds rather just a binary decision 
rule yes/no.

These promising results show that the proposed method is capable to process large amounts
of data, select the most relevant covariates and explain full probability distributions.
This scalable method promises that distributional regression can be applied to large
data sets such as satellite observations (for new developments see, e.g., \citealp{holmlund2021}),
which will enable better descriptions of flash rates across Europe and Africa
(for a recent climatology see, e.g., \citealp{chakraborty2022}).

\section{Summary} \label{sec:summary}

This paper presents a novel algorithm for batchwise backfitting with structured additive
distributional regression models, which is applicable to a much broader class of models as
compared to the approach of \citet{sdr:Li+Wood:2019}. The algorithm combines traditional backfitting
with the ideas of SGD algorithms developed for very large data sets. It converges extremely fast due
to an adaptive learning rate vector employing readily available unbiased estimates of the Hessian, similar to natural gradients.
In combination with the flat file data format, it is thus possible to estimate virtually arbitrarily
large models on a conventional laptop, e.g., with $10^7$ observations and more.

Moreover, depending on the hyperparameter settings, smoothing parameter and variable selection
is performed on-the-fly without requiring further computations on additional validation data. This is, to the best of our knowledge, novel and has never been
presented before in structured additive distributional regression. Besides an extensive
simulation study, the advantages of the new algorithm are demonstrated using complex
distributional regression models on a huge data set for lightning count prediction.

In terms of extensions to the presented framework, the confidence intervals that are not yet 
available should be mentioned. Therefore, for the future we plan to extend the algorithm towards 
Bayesian estimation.

\section*{Acknowledgments}

\begin{leftbar}
This project was partially funded by the Austrian Science Fund~(FWF) grant number~$33941$,
and FWF grant number~$31836$ (Thorsten Simon).
We are grateful for data support by Gerhard Diendorfer and Wolfgang Schulz from OVE-ALDIS.
The computational results presented here have been achieved (in part) using the
LEO HPC infrastructure of the University of Innsbruck. Nadja Klein was supported by
the Deutsche Forschungsgemeinschaft (DFG, German Research Foundation)
through the Emmy Noether grant KL 3037/1-1.
\end{leftbar}


\bibliography{sdr}


\newpage

\begin{appendix}

\section{Batchwise Backfitting in bamlss} \label{appendix:estimation}

This section provides introductory examples on how to fit distributional regression
models for very large data sets with the \pkg{bamlss} package and the new batchwise
backfitting algorithm.

After loading the package with
\begin{Schunk}
\begin{Sinput}
R> library("bamlss")
\end{Sinput}
\end{Schunk}
we simulate a data set with $10^7$ observations using the \fct{GAMart} function with
\begin{Schunk}
\begin{Sinput}
R> set.seed(123)
R> d <- GAMart(n = 1e+07, sd = -1)
\end{Sinput}
\end{Schunk}
Then we save the data as a \code{.csv} file and remove the \proglang{R} data frame from
the global environment.
\begin{Schunk}
\begin{Sinput}
R> write.csv(d, file = "d.csv", row.names = FALSE)
R> rm(d)
\end{Sinput}
\end{Schunk}
To design the scenario very realistic with respect to a very large data set, we read the data back
into \proglang{R} as a flat file data frame.
\begin{Schunk}
\begin{Sinput}
R> library("ff")
R> dff <- read.csv.ffdf(file = "d.csv", header = TRUE)
\end{Sinput}
\end{Schunk}

The \fct{GAMart} function with \code{sd = -1} simulates Gaussian data with
$\texttt{y} \sim N(\mu = \eta_{\mu}, \log(\sigma) = \eta_{\sigma})$
and predictors given by
\begin{eqnarray*}
\eta_{\mu} &=& f_1(\texttt{x1}) + f_2(\texttt{x3}) + f_{2d}(\texttt{lon}, \texttt{lat}) \\
\eta_{\sigma} &=& f_3(\texttt{x2}) + f_2(\texttt{x3}) +  f_4(\texttt{x4}).
\end{eqnarray*}
Here functions $f_1(\cdot), \ldots, f_4(\cdot)$ represent univariate smooth functions and function
$f_{2d}(\cdot)$ a smooth two dimensional effect of coordinates \code{lon} and \code{lat}.
Note that the data set contains additional noise variables \code{x5} and \code{x6}.

\subsection{Boosting Variant} \label{sec:boosting}

\begin{figure}[t!]
\centering
\includegraphics[width=1\textwidth]{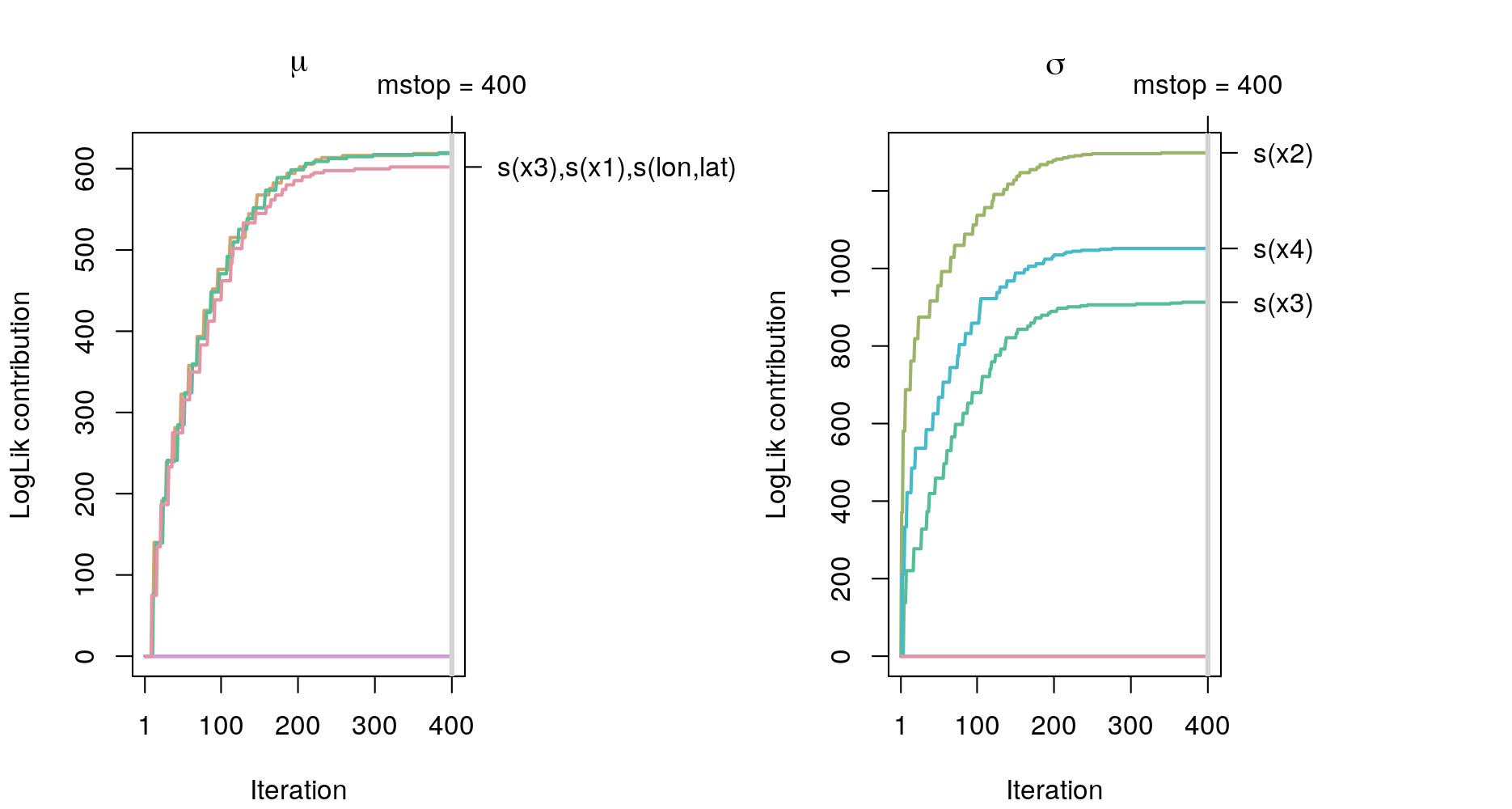}
\caption{\label{fig:contribpaths} Log-likelihood contribution paths for
  selected terms of the Gaussian model.}
\end{figure}

We first illustrate the usage of the new model fitting engine using the boosting variant of
the batchwise backfitting algorithm, see Section~\ref{sec:bbfit}. Therefore, we set up a list of
model formulae, where each formula includes all covariates.
\begin{Schunk}
\begin{Sinput}
R> f <- ~ s(x1) + s(x2) + s(x3) + s(x4) + s(x5) + s(x6) + s(lon, lat)
R> f <- list(update(f, y ~ .), f)
\end{Sinput}
\end{Schunk}
Before estimation can be started, batch indices can be specified with
\begin{Schunk}
\begin{Sinput}
R> set.seed(456)
R> n <- nrow(dff)
R> batch_ids <- lapply(1:400, function(...) sample(n, size = 10000))
\end{Sinput}
\end{Schunk}
I.e., we use 400 batches with batchsize of $10000$ such that the algorithm can practically see
the entire data once. Then, the model is estimated with
\begin{Schunk}
\begin{Sinput}
R> b <- bamlss(f, data = dff, family = "gaussian",
+    sampler = FALSE, optimizer = opt_bbfit,
+    nu = 0.1, always = FALSE, AIC = TRUE, eps_loglik = 0.0001,
+    batch_ids = batch_ids, select = TRUE,
+    ff_name = "ff_simdata", delete = FALSE, overwrite = FALSE,
+    light = TRUE)
\end{Sinput}
\end{Schunk}
Note that the data frame \code{dff} is an \code{"ffdf"} data frame. For processing all the
design matrices for estimating the model, we therefore
specify a directory name with \code{ff\_name = "ff\_simdata"}, where all matrices can
be stored as \pkg{ff} objects. The directory \code{ff\_simdata} is created in the current
working directory and will not be deleted after estimation if \code{delete = FALSE}. This
has the advantage, that the \pkg{ff} matrices can be reused for other models, e.g., using
a different distribution, i.e., setting up more models will be much faster. If
\code{overwrite = FALSE}, the directory for storing the \pkg{ff} objects will not be overwritten
when calling \fct{bamlss}. Moreover, option \code{light = TRUE} can be used to reduce
the memory footprint of the final returned object even more.
Here, we need to set \code{sampler = FALSE} in order to switch off subsequent MCMC sampling.
The \fct{opt\_bbfit} optimizer function is used as the model fitting engine, for which we
specify the step length control parameter \code{nu = 0.1}. Argument \code{always = FALSE} causes
model terms to be updated only if the relative improvement of the \textit{out-of-sample}
log-likelihood (on the next batch) is larger than \code{0.0001} (controlled by argument
\code{eps\_loglik}). By setting \code{AIC = TRUE} the smoothing variances are selected by the
\textit{out-of-sample} AIC, otherwise the \textit{out-of-sample} log-likelihood is used.
As we are interested in the boosting variant of the batchwise backfitting
algorithm in this case, we need to set argument \code{select = TRUE}, i.e., only the model term
with the largest contribution to the log-likelihood in the next batch will be updated.
On a Linux system with Intel(R) Core(TM) i7-8550U CPU \@ 1.80GHz processors, the estimation time
is about 55 minutes, which is considerably fast for such a large data set.
Selection frequencies of the 400 boosting iterations and individual log-likelihood contributions
can be shown with
\begin{Schunk}
\begin{Sinput}
R> contribplot(b)
\end{Sinput}
\end{Schunk}
\begin{Schunk}
\begin{Soutput}
mu
             Sel. freq.
s.s(x1)           0.116
s.s(x3)           0.116
s.s(lon,lat)      0.112
p                 0.004
s.s(x2)           0.000
s.s(x4)           0.000
s.s(x5)           0.000
s.s(x6)           0.000

sigma
             Sel. freq.
p                 0.220
s.s(x3)           0.156
s.s(x4)           0.140
s.s(x2)           0.136
s.s(x1)           0.000
s.s(x5)           0.000
s.s(x6)           0.000
s.s(lon,lat)      0.000
\end{Soutput}
\end{Schunk}
The selection frequencies reveal that the boosting variant of the batchwise backfitting algorithm
selected the correct model terms, the corresponding contribution paths are shown in
Figure~\ref{fig:contribpaths}. The paths also show the convergence of the algorithm already
at about iteration 200.

\subsection{Resampling}  \label{sec:ssampling}

This section demonstrates the \textit{resampling} variant with slice sampling of
smoothing variances of the batchwise backfitting algorithm. From the selected model terms
we set up a new formula
\begin{Schunk}
\begin{Sinput}
R> f <- list(
+    y ~ s(x1) + s(x3) + s(lon, lat),
+      ~ s(x2) + s(x3) + s(x4)
+  )
\end{Sinput}
\end{Schunk}
and only slightly modify the arguments supplied to the main model fitting function \fct{bamlss}
\begin{Schunk}
\begin{Sinput}
R> m <- bamlss(f, data = dff, family = "gaussian",
+    sampler = FALSE, optimizer = opt_bbfitp,
+    AIC = TRUE, slice = TRUE, batch_ids = batch_ids,
+    ff_name = "ff_simdata", delete = FALSE, overwrite = FALSE,
+    light = TRUE)
\end{Sinput}
\end{Schunk}
Estimation takes about 24 minuts. Note that we use a wrapper version \fct{opt\_bbfitp} for
estimating the model. The only difference is that the parameters are stored as
\code{"mcmc"} ``samples'' in the returned object,
so all extractor functions such as \fct{predict}, \fct{residuals}, etc., can be used
similarly to estimating full Bayesian models with MCMC. For details on using \pkg{bamlss}, see
\citet{sdr:Umlauf+Klein+Simon+Zeileis:2021} and the project website \url{http://bamlss.org/}.
A major advantage of the new infrastructures, including the \pkg{ff} package, is that all design
matrices can be reused and do not need to be recomputed, saving large amounts of runtime. To use
this feature, all that is required is careful handling of the \code{ff\_name}, \code{delete} and
\code{overwrite} arguments (use as described in the last section). By setting \code{slice = TRUE},
slice sampling of smoothing parameters in combination with the resampling variant with
\code{nu = 1}, \code{eps\_loglik = -Inf} and \code{always = TRUE} is
used for batchwise backfitting, i.e., updates are always accepted. Convergence of the algorithm
can be inspected by, e.g., coefficient paths of the parameters.

\newpage

\section{Simulation Results} \label{appendix:simresults}

In this section we show all the results of the simulation study presented in Section~\ref{sec:simulation}.

\begin{figure}[ht!]
\centering
\includegraphics[width=0.9\textwidth]{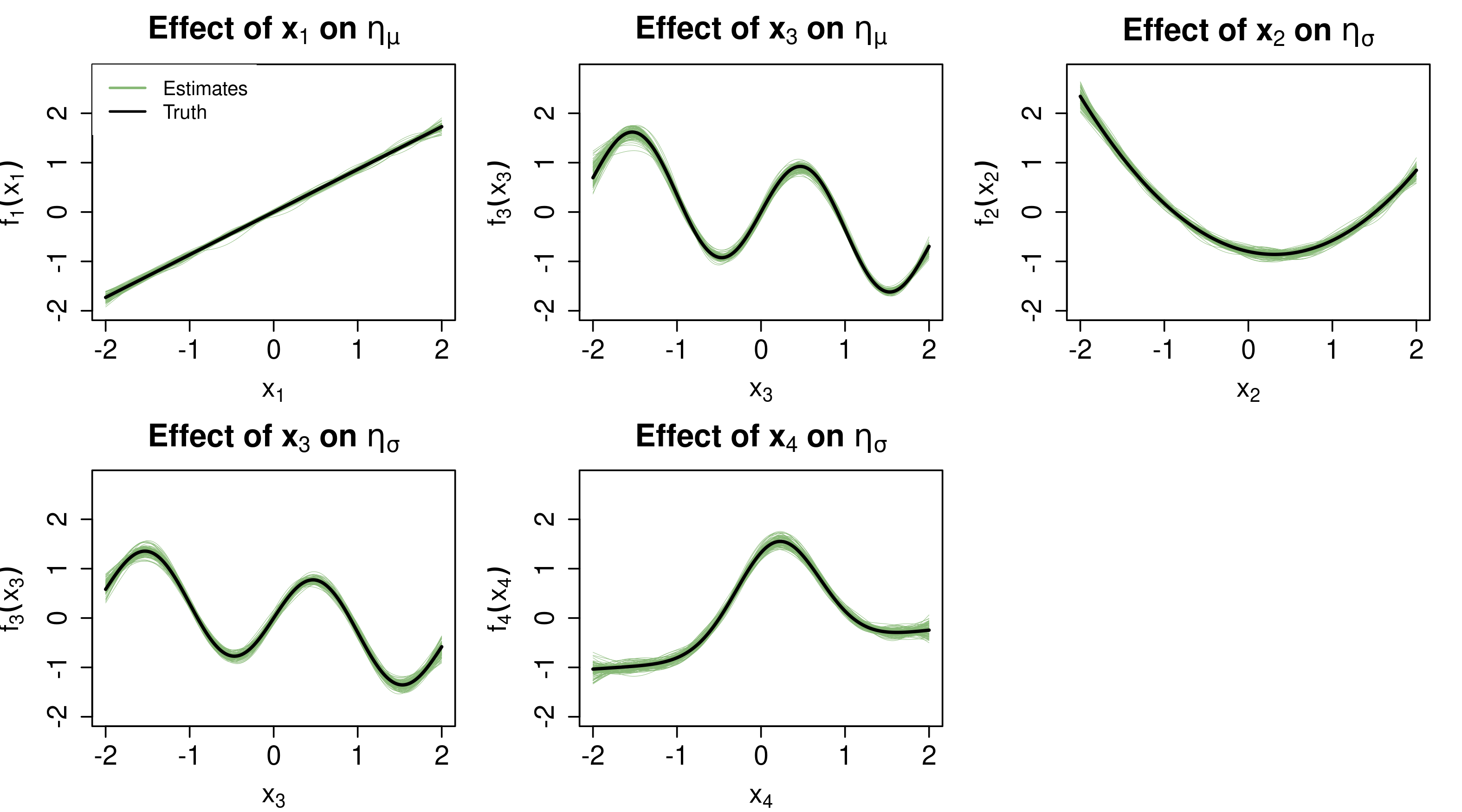}
\caption{\label{fig:NO_effects1} Functions used in the simulation study. Green lines correspond
  to the estimated effects of the \code{opt\_bbfit} model of the 100 replications in the
  simulation setting: normal distribution \code{NO}, observations \code{n = 5000}, noise variables
  \code{nnoise = 10} and correlation \code{rho = 0}. }
\end{figure}

\begin{figure}[ht!]
\centering
\includegraphics[width=0.71\textwidth]{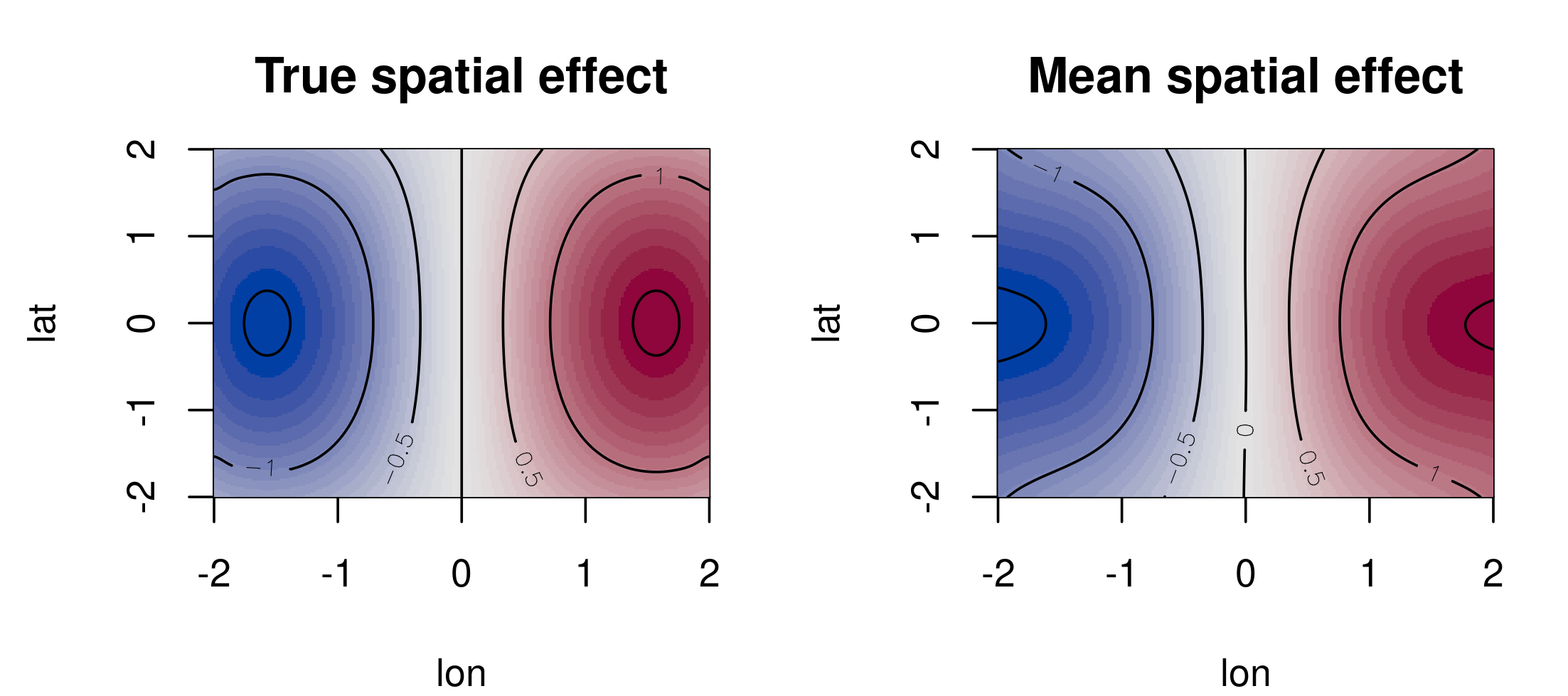}
\caption{\label{fig:NO_effects1} Spatial effect used in the simulation study. Spatial effects correspond to the estimated two-dimensional effects of the method \code{opt\_bbfit} in the 100 replications of the  simulation setting: normal distribution \code{NO}, observations \code{n = 5000}, noise variables
  \code{nnoise = 10} and correlation \code{rho = 0}.  The mean was calculated over the 100 different estimates of the two-dimensional effect. }
\end{figure}

\begin{figure}[ht!]
\centering
\includegraphics[width=0.71\textwidth]{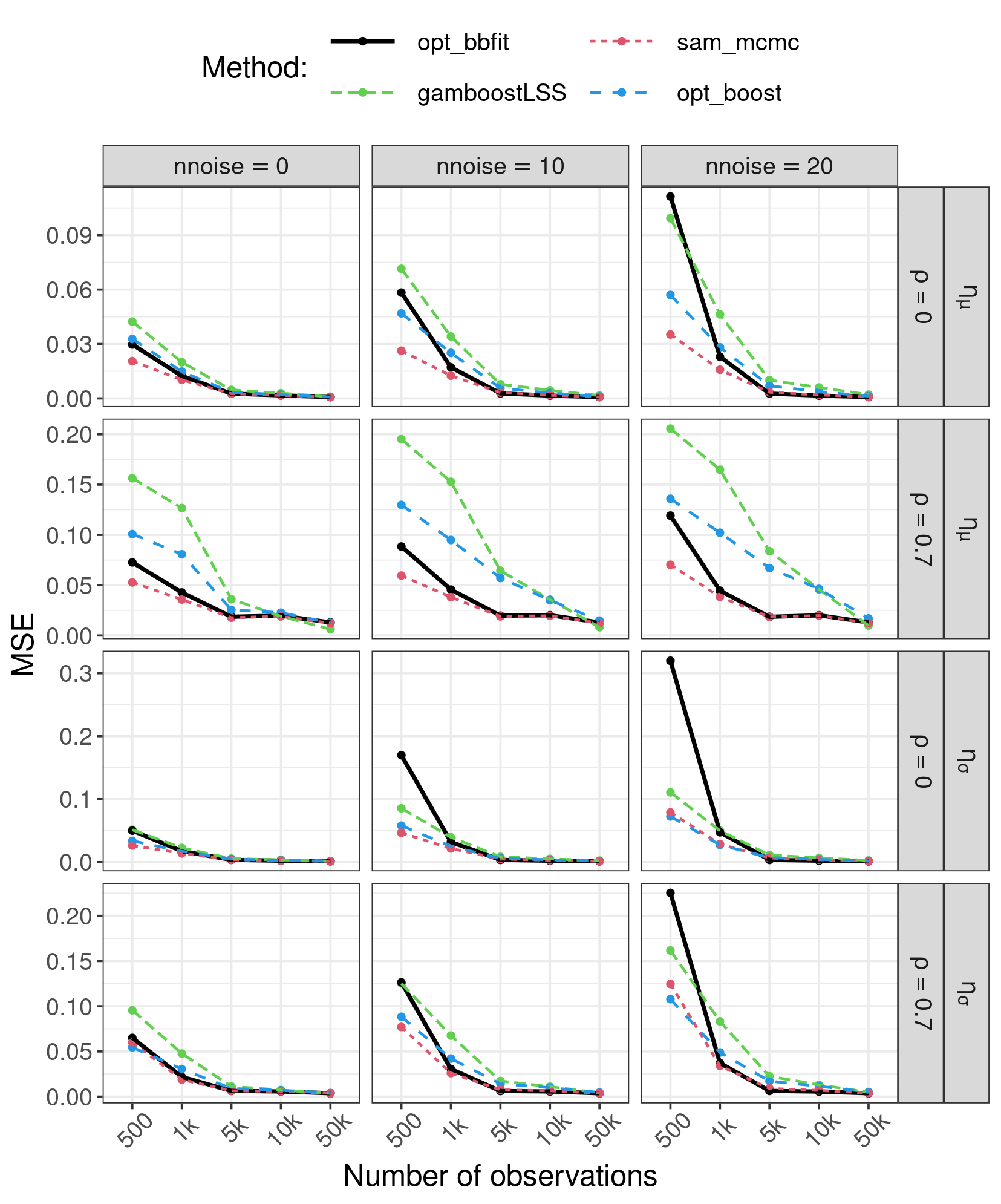}
\caption{\label{fig:resGA} Simulation study. Average MSE with the \code{GA} distribution for
  predictor $\eta_{\mu}$ and $\eta_{\sigma}$ for different number of observations,
  correlation and noise variable settings.}
\end{figure}

\begin{figure}[ht!]
\centering
\includegraphics[width=0.71\textwidth]{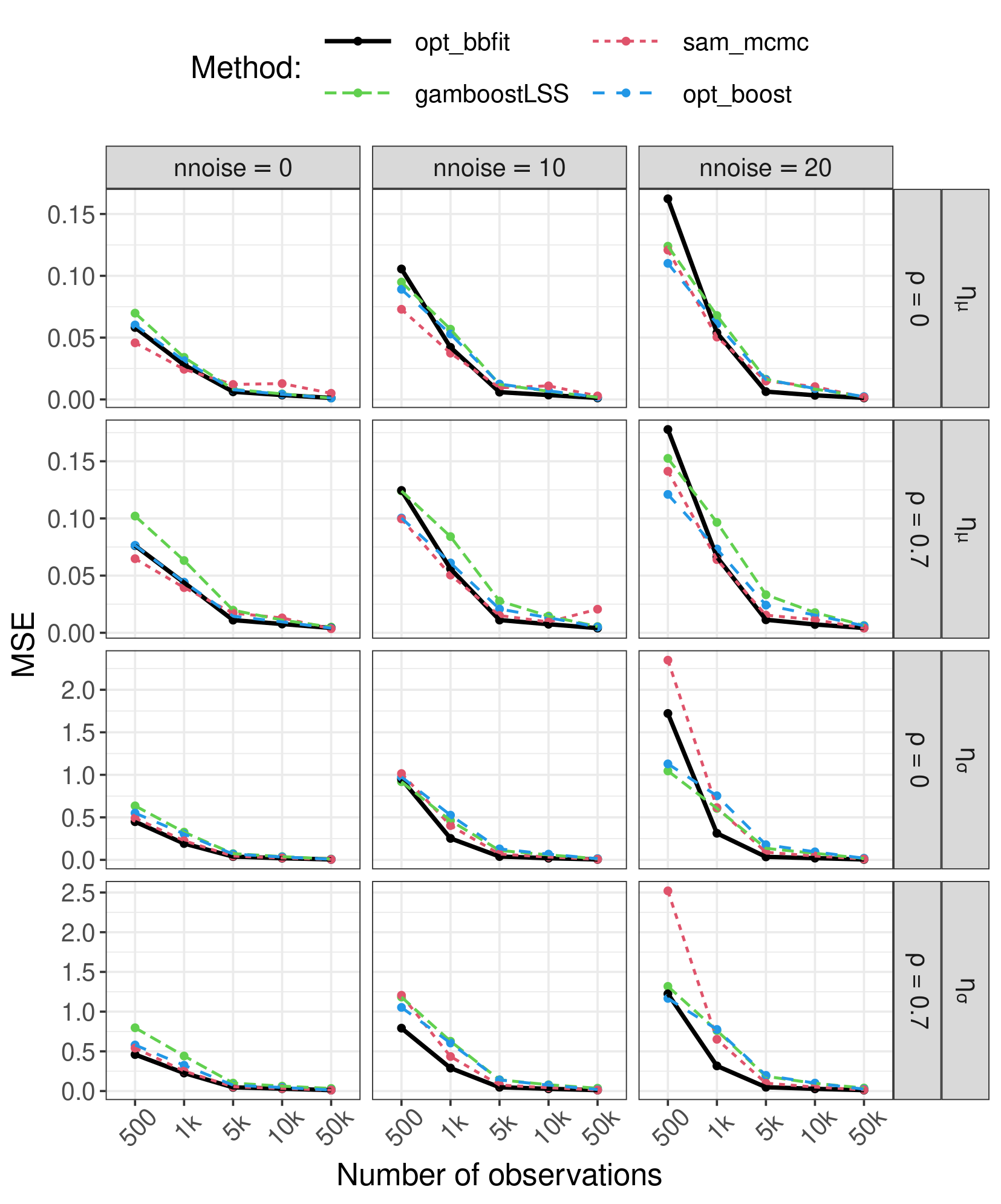}
\caption{\label{fig:resZAP} Simulation study. Average MSE with the \code{ZAP} distribution for
  predictor $\eta_{\mu}$ and $\eta_{\sigma}$ for different number of observations,
  correlation and noise variable settings.}
\end{figure}

\begin{figure}[ht!]
\centering
\includegraphics[width=0.71\textwidth]{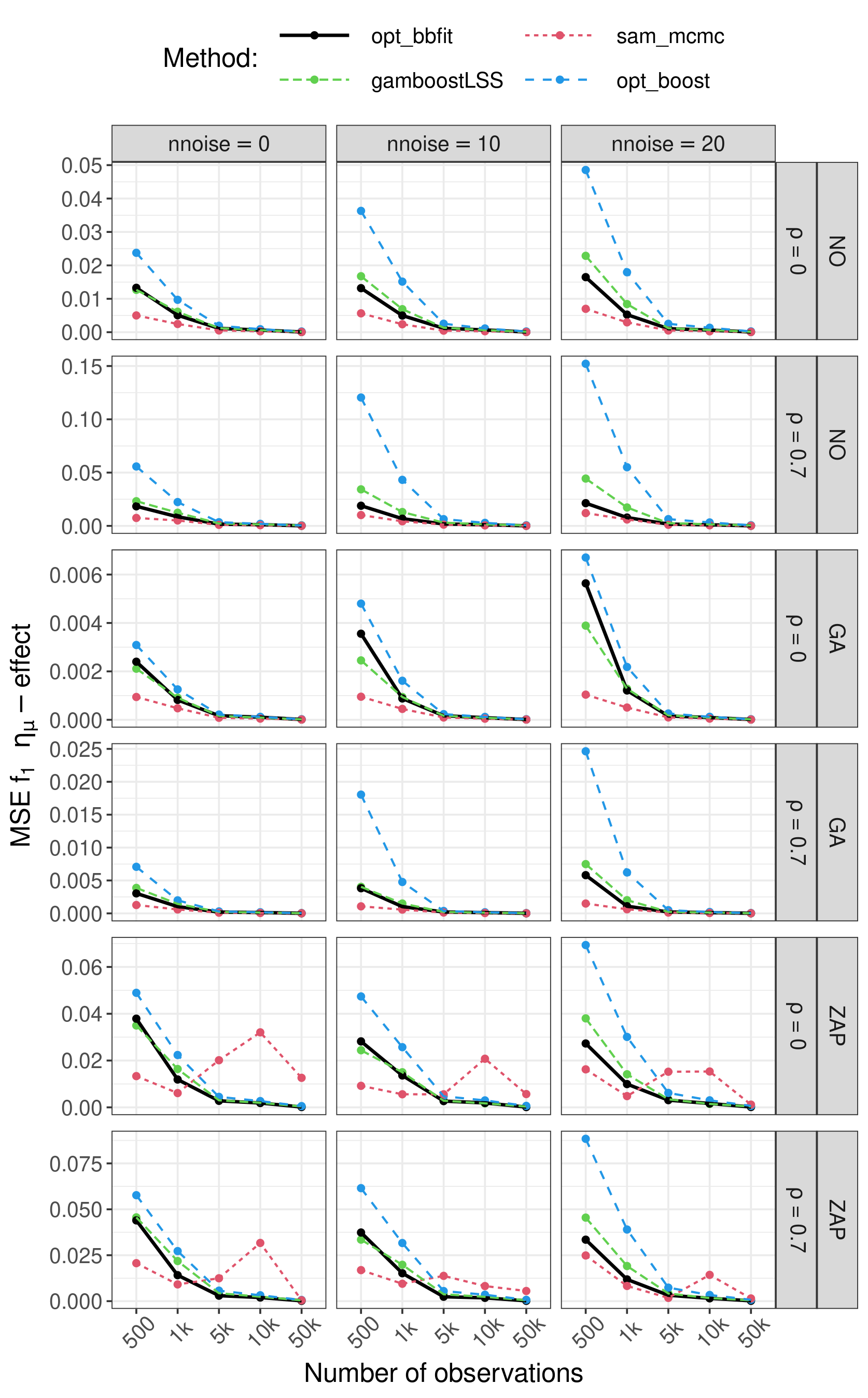}
\caption{\label{fig:f1}  Simulation study. Average MSE of $f_1$ effect of $\eta_{\mu}$ for all distributions,
  different number of observations, correlation and noise variables.}
\end{figure}

\begin{figure}[ht!]
\centering
\includegraphics[width=0.71\textwidth]{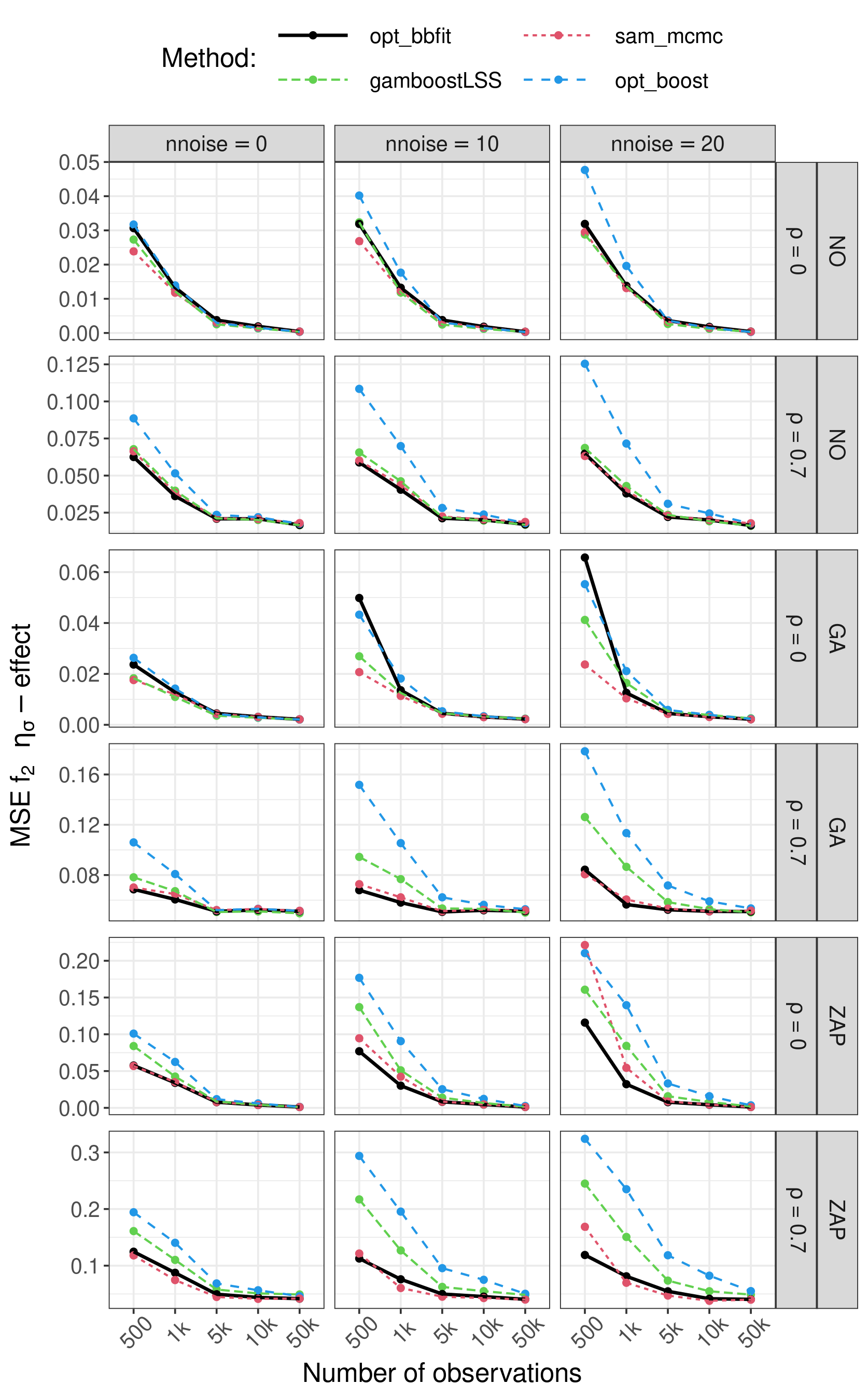}
\caption{\label{fig:f2s}  Simulation study. Average MSE of $f_2$ effect of $\eta_{\sigma}$ for all distributions,
  different number of observations, correlation and noise variables.}
\end{figure}

\begin{figure}[ht!]
\centering
\includegraphics[width=0.71\textwidth]{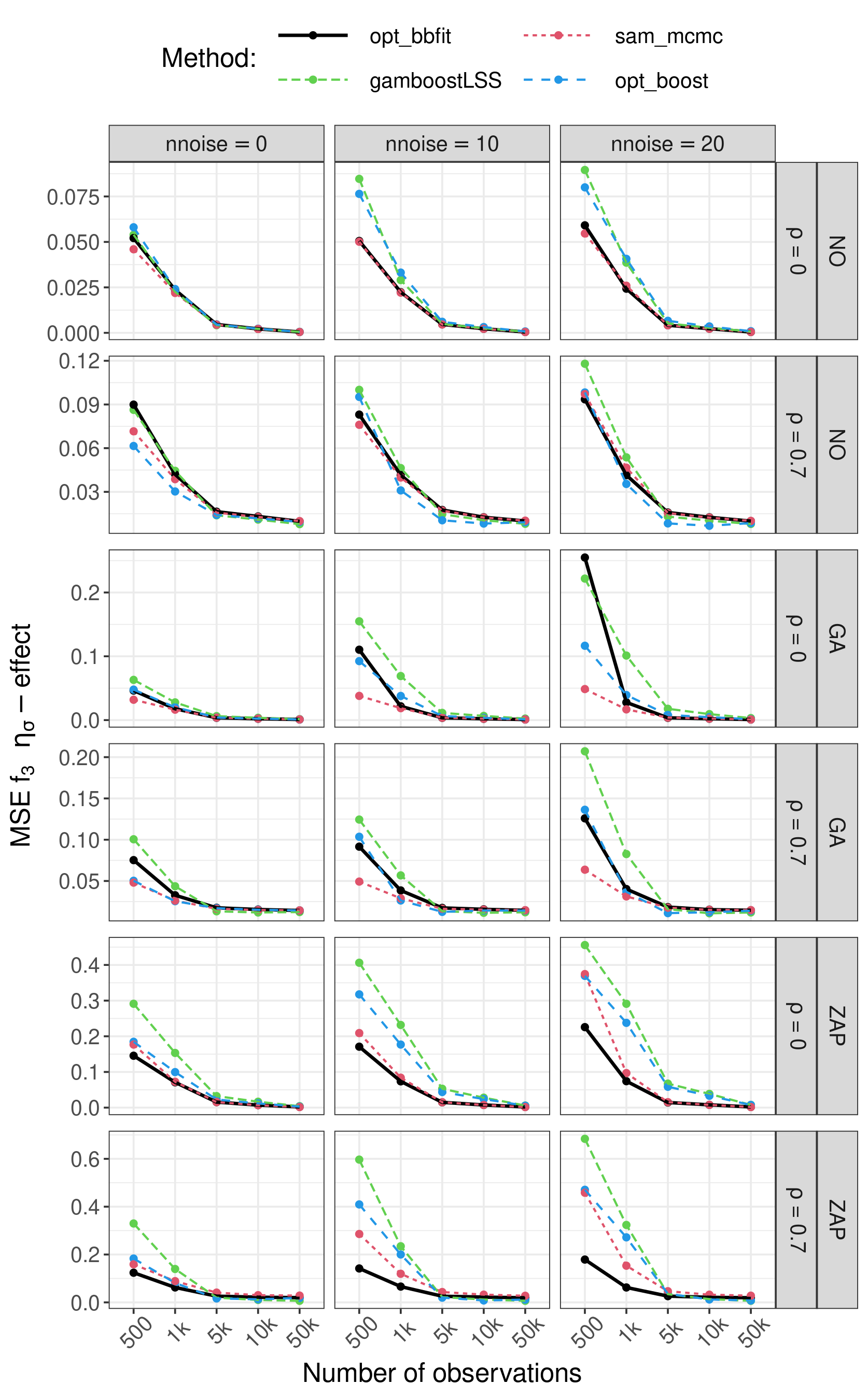}
\caption{\label{fig:f3s}  Simulation study. Average MSE of $f_3$ effect of $\eta_{\sigma}$ for all distributions,
  different number of observations, correlation and noise variables.}
\end{figure}

\begin{figure}[ht!]
\centering
\includegraphics[width=0.71\textwidth]{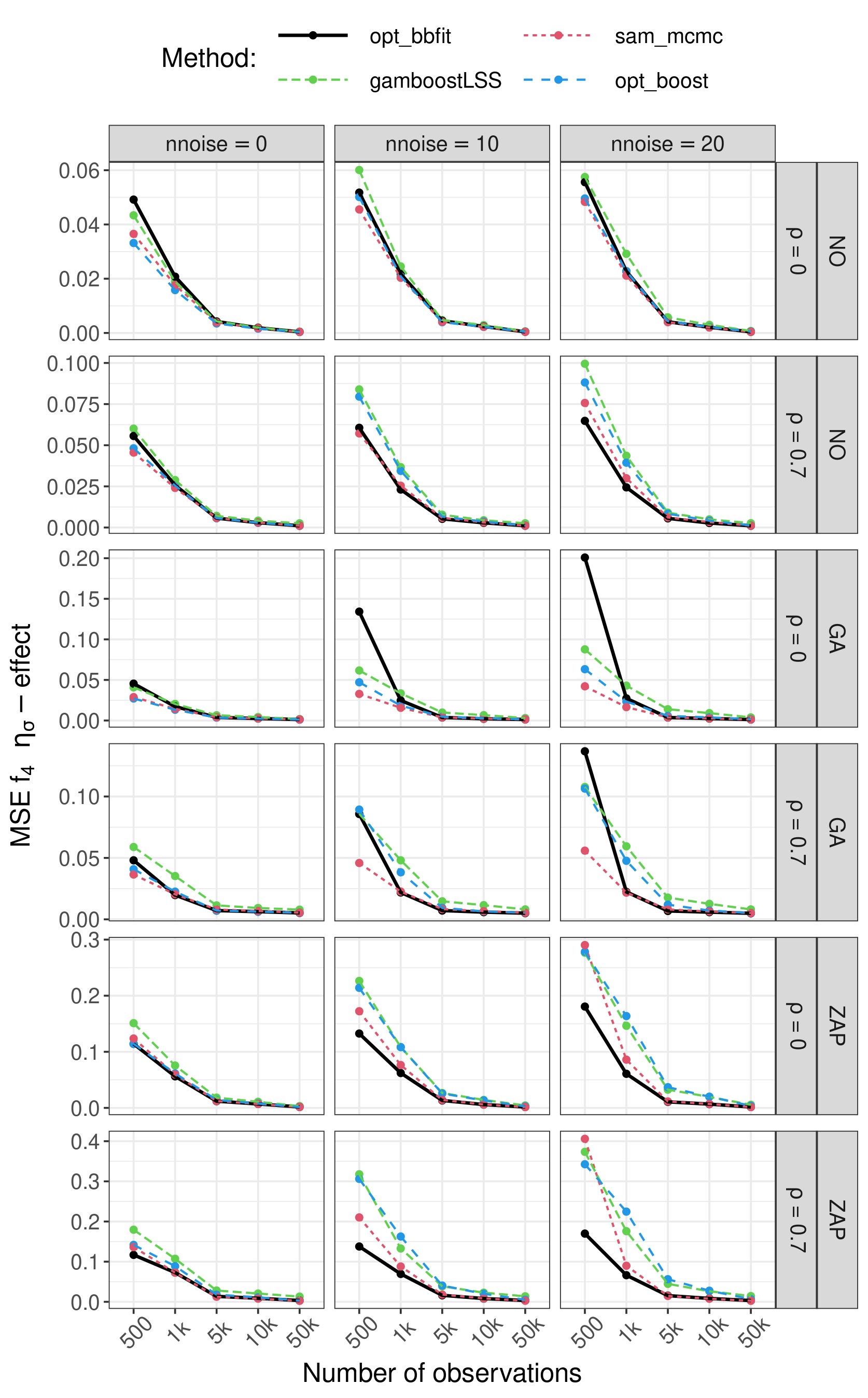}
\caption{\label{fig:f4s}  Simulation study. Average MSE of $f_4$ effect of $\eta_{\sigma}$ for all distributions,
  different number of observations, correlation and noise variables.}
\end{figure}

\begin{figure}[ht!]
\centering
\includegraphics[width=0.855\textwidth]{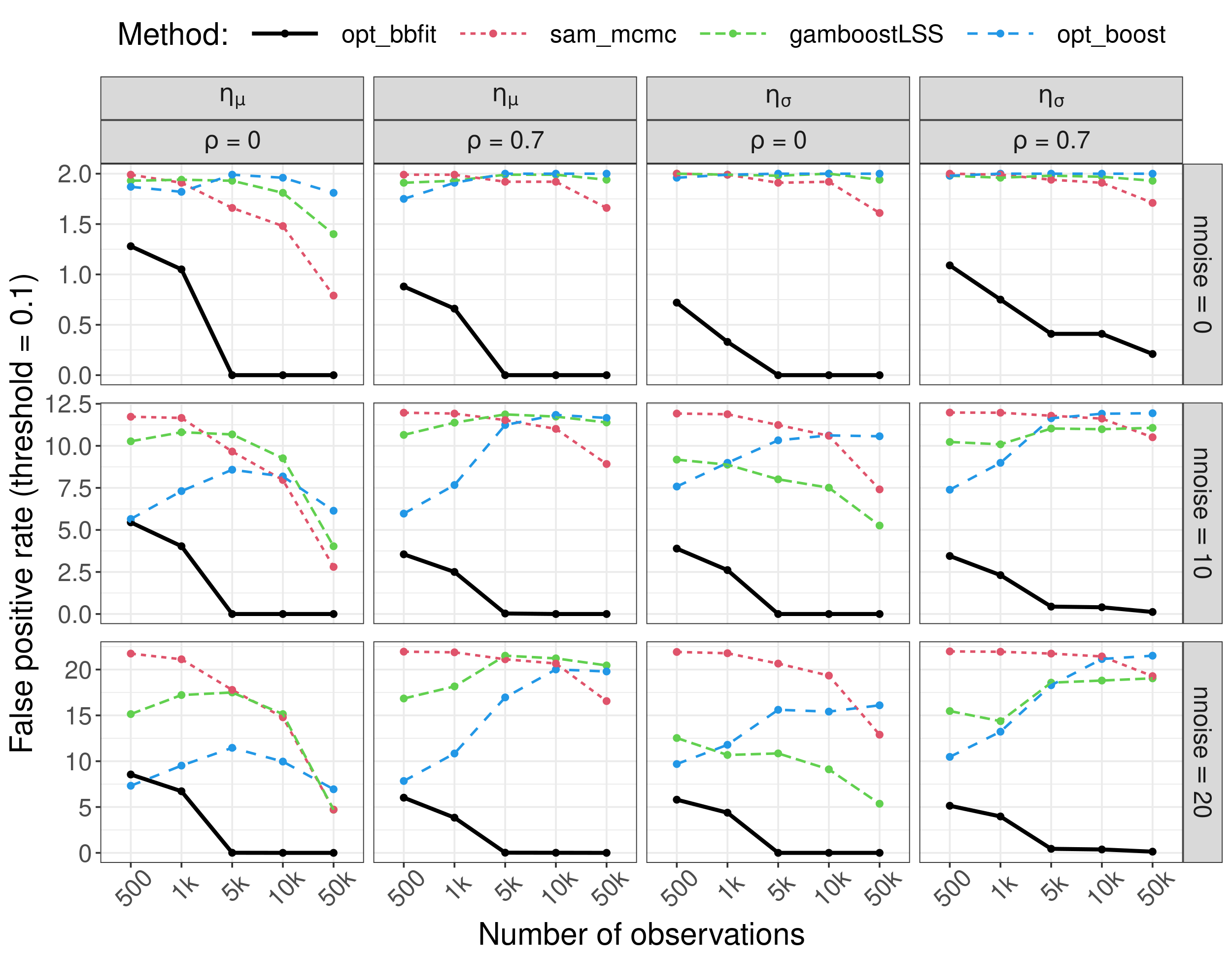}
\caption{\label{fig:resGA_fp} Simulation study. Average false positive rate of the \code{GA} distribution
  for predictor $\eta_{\mu}$ and $\eta_{\sigma}$ for different number of observations,
  correlation and noise variable settings.}
\end{figure}

\begin{figure}[ht!]
\centering
\includegraphics[width=0.855\textwidth]{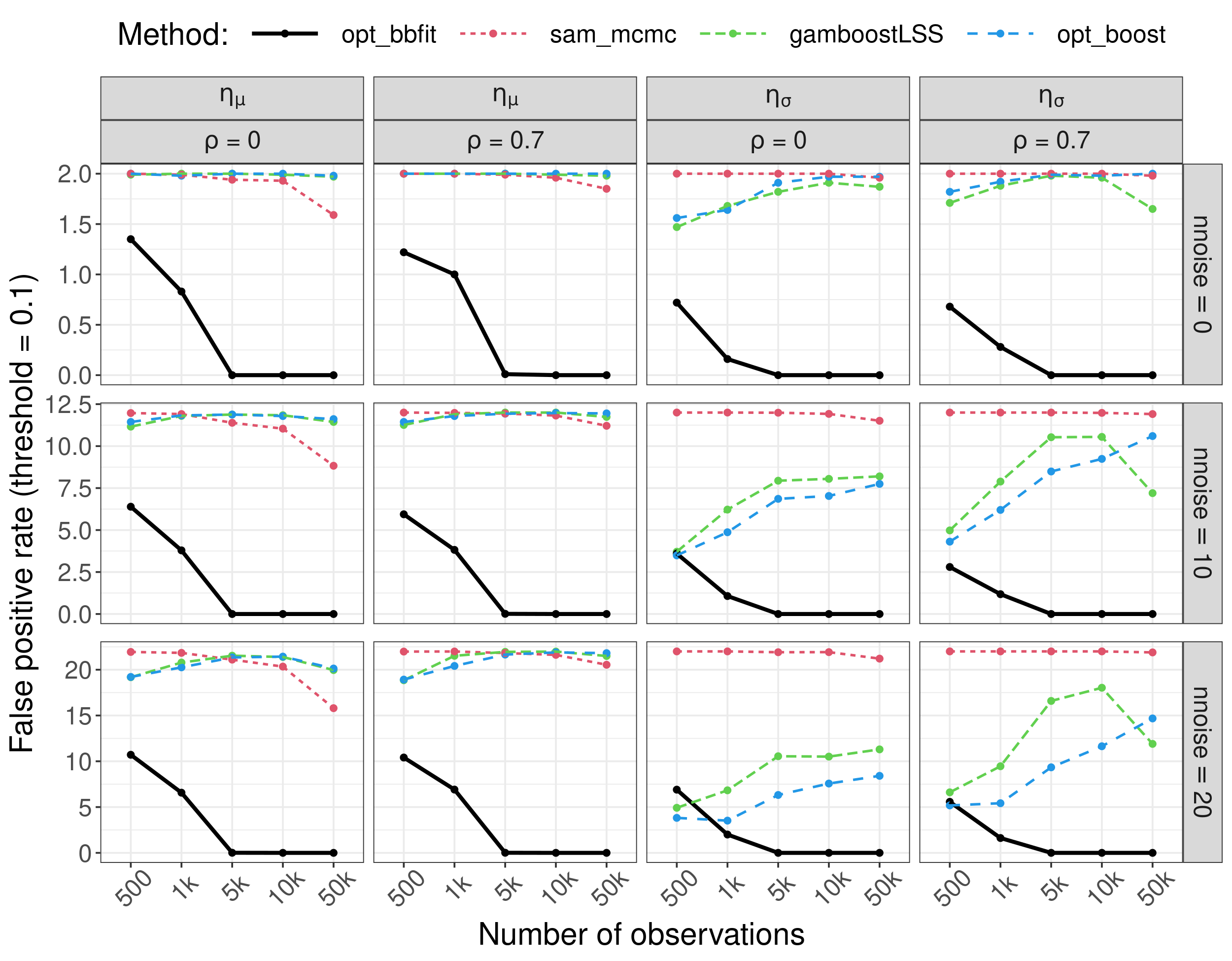}
\caption{\label{fig:resZAP_fp} Simulation study. Average false positive rate of the \code{ZAP} distribution
  for predictor $\eta_{\mu}$ and $\eta_{\sigma}$ for different number of observations,
  correlation and noise variable settings.}
\end{figure}

\begin{figure}[ht!]
\centering
\includegraphics[width=0.9\textwidth]{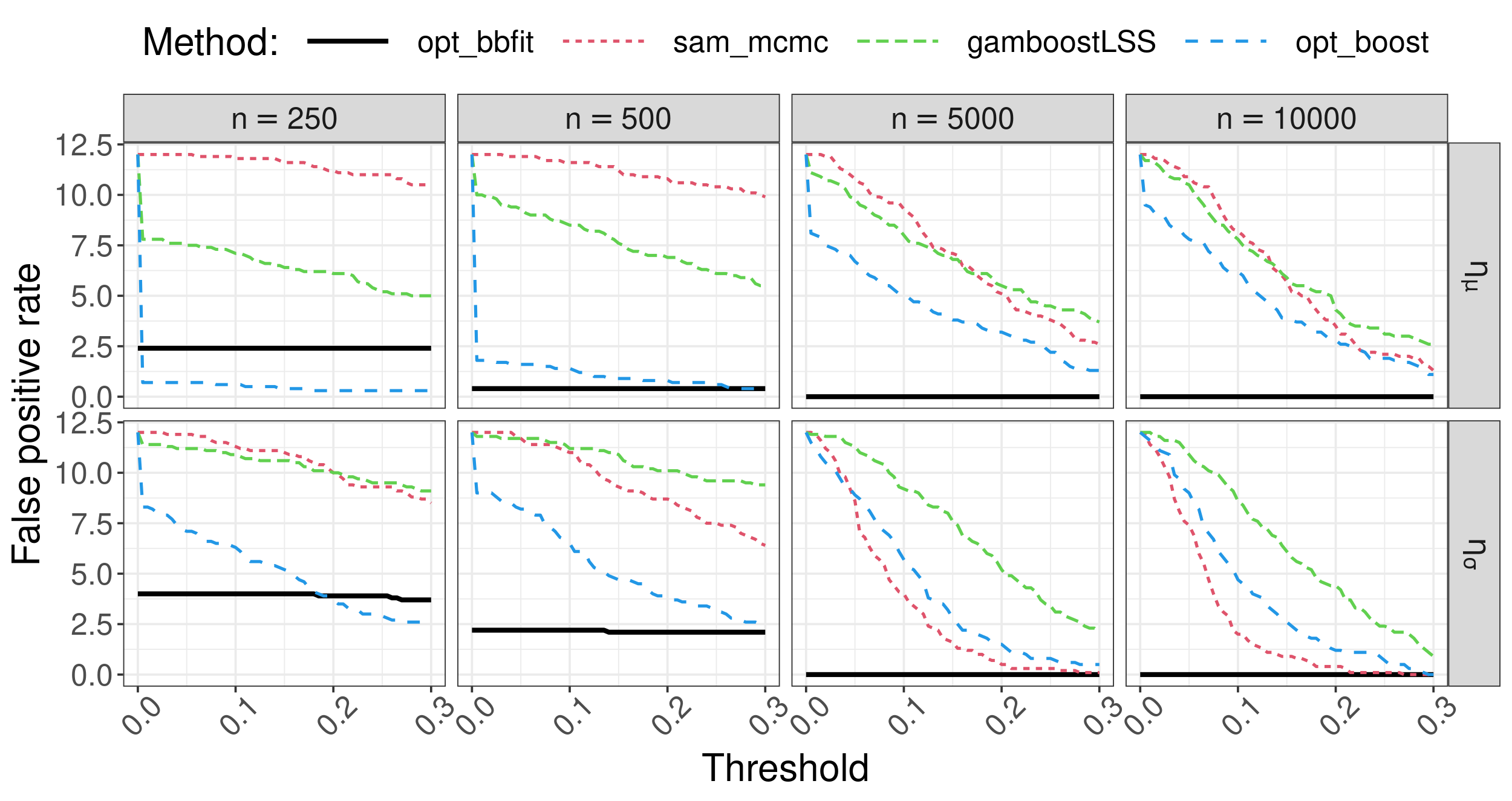}
\caption{\label{fig:resthres} Simulation study. Average false positive rate of $10$ replications per
  setting for different threshold values. Setting: Number of observations varies, \code{NO},
  $\rho = 0.7$ and \code{nnoise = 10}. For $n \geq 500$ the method \code{opt\_bbfit} is the
  best in terms of excluding non-informative variables.}
\end{figure}

\clearpage

\end{appendix}


\end{document}